\documentclass[pre,tighten,eqsecnum,showpacs]{revtex4}

\usepackage{amssymb,amsmath}

\usepackage{graphicx}
\usepackage{color}

\definecolor{dark-green}{rgb}{0.2,0.6,0.3}

\newcommand\beq{\begin{equation}}
\newcommand\eeq{\end{equation}}
\newcommand\beqa{\begin{eqnarray}}
\newcommand\eeqa{\end{eqnarray}}
\newcommand{\dd}{\text{d}}

\newcommand{\al}{\alpha}

\newcommand{\nuzt}{\nu_{0|2}}
\newcommand{\nuto}{\nu_{2|1}}

\newcommand{\nufz}{\nu_{4|0}}

\begin{document}

\title{Navier-Stokes transport coefficients for driven inelastic Maxwell models}
\author{Mois\'es G. Chamorro}
\email{moises@unex.es}
\author{Vicente Garz\'o}
\email{vicenteg@unex.es} \homepage{http://www.eweb.unex.es/eweb/fisteor/vicente/}
\affiliation{Departamento de F\'{\i}sica, Universidad de Extremadura, E-06071 Badajoz, Spain}
\author{Francisco Vega Reyes}
\email{fvega@unex.es} \homepage{http://www.unex.es/eweb/fisteor/fran/}
\affiliation{Departamento de F\'{\i}sica, Universidad de Extremadura, E-06071 Badajoz, Spain}

 \begin{abstract}
We calculate in this work the Navier-Stokes transport coefficients from the Boltzmann equation for $d$-dimensional inelastic Maxwell models. By granular gas we mean here a low density system of identical spheres that lose a fraction of their kinetic energy after collisions. In the present work, the granular gas is fluidized by the presence of a thermostat that aides the system to reach a steady state. The thermostat is composed by two terms: a random force and a drag force. The combined action of both forces, that act homogeneously on the granular gas, tries to mimic the interaction of the set of particles with a surrounding fluid. The Chapman-Enskog method is applied to solve the inelastic Boltzmann equation to first order in the deviations of the hydrodynamic fields from their values in the homogeneous steady state. Since the collisional cooling cannot be compensated locally for by the heat produced by the driving forces, the reference (zeroth-order) distribution function $f^{(0)}$ depends on time through its dependence on the granular temperature. To simplify the analysis and obtain explicit forms for the transport coefficients, the steady state conditions are considered. A comparison with previous results obtained for inelastic hard spheres is also carried out.


\end{abstract}

\draft
\date{\today}
\maketitle

\section{Introduction}
\label{sec1}

Granular matter (systems composed of many  mesoscopic particles) under rapid flow conditions can be modeled as a ``granular gas'', namely a gas of hard spheres dissipating part of their kinetic energy during binary collisions (inelastic hard spheres, IHS). In the simplest model, the spheres are completely smooth and the degree of inelasticity is characterized by the so-called coefficient of normal restitution $\al \leq 1$, that is assumed to be \emph{constant}. At the level of kinetic theory, all the relevant information on the state of the gas is provided by the one-particle velocity distribution function $f(\mathbf{r},\mathbf{v},t)$. For a low-density gas, the Boltzmann equation has been conveniently modified to account for inelastic binary collisions \cite{G02,BP04} and the corresponding Navier-Stokes transport coefficients \cite{NS} for states with small spatial gradients have been obtained by means of the Chapman-Enskog expansion \cite{CC70} around the \emph{local} version of the homogeneous cooling state. Similarly to the case of a gas with elastic collisions, the exact form of the Navier-Stokes transport coefficients is not known since they are given in terms of the solutions of a coupled set of linear integral equations. A good approach to the exact form of these coefficients can be obtained by considering the leading terms in a Sonine polynomial expansion of the distribution function $f(\mathbf{r},\mathbf{v},t)$ \cite{CC70}. Despite this approach, the theoretical predictions compare in general quite well with computer simulations even for relatively small values of the coefficient of restitution $\al$ \cite{DSMC}.

On the other hand, due to collisional kinetic energy loss, an additional source of  energy is needed in order to keep the system under rapid flow and reach a steady state. This external energy can be supplied to the system from the boundaries (for instance, from vibrating walls \cite{YHCMW02}) or by bulk driving as in air-fluidized beds \cite{AD06,SGS05}. Under certain experimental conditions, the bulk driving is homogeneous, and this is the case we consider in this work. In fact, it is quite usual in computer simulations to homogeneously heat the system by the action of an external driving force \cite{andrea,emmanuel}. This type of external forces are called ``thermostats'' \cite{EM90}. Although thermostats have been widely used in the past to study granular flows, their influence on the dynamic properties of the system (for elastic and granular fluids) is not completely understood yet \cite{DSBR86,GSB90,GS03}.

We will consider in this work that the granular gas is fluidized by a thermostat composed by two different terms: (i) a drag force proportional to the velocity of the particle and (ii) a stochastic force with the form of a Gaussian white noise where the particles are randomly kicked between collisions \cite{WM96}. While the first term attempts to model the friction of grains with a viscous interstitial fluid, the second term models the energy transfer from the surrounding fluid to granular particles. The transport coefficients of the granular gas driven by this combined thermostat have been recently determined \cite{GCV13}.  Like in the undriven case \cite{NS}, the forms of the transport coefficients involve the evaluation of certain collision integrals that cannot be \emph{exactly} computed due to the complex mathematical structure of the (linearized) Boltzmann collision operator for IHS. Thus, in order to get explicit expressions for the above coefficients one has to consider additional approximations. A possible way of circumventing these technical difficulties inherent to IHS, while keeping the structure of the Boltzmann collision operator, is to consider the so-called inelastic Maxwell models (IMM), i.e., models for which the collision rate is independent of the relative velocity of the two colliding particles. In the case of hard spheres with elastic collisions (conventional molecular gases), Maxwell models are characterized by a repulsive potential that (in three dimensions) is proportional to the inverse fourth power of distance between particles. For inelastic collisions, Maxwell models can be introduced in the framework of the Boltzmann equation at the level of the cross section, without any reference to a specific interaction potential \cite{E81}. In addition, apart from its academic interest, it is worthwhile remarking that experiments \cite{KSSAOB05} for magnetic grains with dipolar interactions are well described by IMM.

Therefore, the motivation of the paper is twofold. On the one hand, the knowledge of the first collisional moments
for IMM allows one to re-examine the problem studied in Ref.\ \cite{GCV13} in the context of the (inelastic) Boltzmann equation
and without taking any additional and sometimes uncontrolled approximations. On the other hand, the comparison between
the results obtained from IMM with those derived from IHS \cite{GCV13} can be used again as a test to assess the
reliability of IMM as a prototype model for characterizing real granular flows. Previous comparisons
have shown a mild qualitative agreement in the freely cooling case  \cite{S03,GA05} while the agreement between IMM and IHS significantly increases for low order velocity moments in the case of driven states (for instance, the simple shear flow problem) \cite{G03,G07,SG07,GT10}.

The main advantage of using IMM instead of IHS is that a velocity moment of order $k$ of the Boltzmann collision operator
only involves moments of order less than or equal to $k$. This allows to evaluate the Boltzmann collision moments without
the explicit knowledge of the distribution function \cite{GS07}. This property opens up the search of exact solutions to the
Boltzmann equation and justifies the interest of physicists and mathematicians in IMM in the last years \cite{IMM,GS11}.
Thus, in this paper, we determine the exact forms of the shear viscosity $\eta$, the thermal conductivity $\kappa$
and the transport coefficient $\mu$ (that relates the heat flux with the density gradient) as a function of
the coefficient of restitution $\alpha$ and the thermostat forces intensity. As for IHS \cite{GCV13}, the expressions
of $\eta$, $\kappa$ and $\mu$ are obtained by solving the Boltzmann equation for IMM up to first order in the spatial
gradients by means of the Chapmnan-Enskog expansion \cite{CC70}. A subtle point of the Chapman-Enskog solution derived here is that the zeroth-order distribution $f^{(0)}$ is not in general a stationary distribution since the collisional cooling cannot be compensated \emph{locally} for by the energy supplied by the thermostat \cite{GCV13,GMT13}. Such energy unbalance introduces new contributions to the transport coefficients, which not were considered in previous works \cite{GM02} where local steady state was assumed at zeroth-order.

The plan of the paper is as follows. In section \ref{sec2}, the Boltzmann equation for driven IMM is introduced and the explicit expressions for the second and third-degree collisional moments are given. Section \ref{sec3} deals with the steady homogeneous state where a scaling solution is proposed that depends on granular temperature through two dimensionless parameters (dimensionless velocity and reduced noise strength) \cite{CVG13}. Section \ref{sec4} addresses the Chapman-Enskog expansion around the \emph{unsteady} reference distribution $f^{(0)}(\mathbf{r},\mathbf{v},t)$ while the Navier-Stokes transport coefficients are obtained in section \ref{sec5}. The explicit dependence of $\eta$, $\kappa$ and $\mu$ on the parameters of the system requires in general  to solve numerically a set of nonlinear differential equations. As for IHS \cite{GCV13}, those differential equations become simple algebraic equations when the steady state conditions are considered. The dependence of the transport coefficients on the coefficient of restitution is illustrated and compared with the results for IHS \cite{GCV13} in section \ref{sec5}. The comparison shows in general a good qualitative agreement, although quantitative discrepancies between both interaction models appear as inelasticity increases. The paper is closed in section \ref{sec6} with a brief discussion of the results derived in this paper.

\section{Inelastic Maxwell models}
\label{sec2}

Let us consider a granular fluid modeled as Maxwell gas of inelastic particles. Inelasticity in the translational degrees of freedom of the grains is measured by a {\em constant} and positive coefficient of restitution $\alpha \leq 1$. As said in the Introduction, the granular gas is driven by two different external \emph{nonconservative} forces: (i) a stochastic force where the particles are randomly kicked between collisions \cite{WM96} and (ii) a viscous drag force which mimics  the interaction of the grains with an effective background ``bath''. Under these conditions, the one-particle velocity distribution function $f({\bf r}, {\bf v}, t)$ obeys the \emph{inelastic} Boltzmann equation
\begin{equation}
\label{2.1}
\frac{\partial f}{\partial t}+{\bf v}\cdot \nabla
f-\frac{\gamma_\text{b}}{m} \frac{\partial}{\partial
{\bf v}}\cdot {\bf V} f-\frac{1}{2}\xi_\text{b}^2\frac{\partial^2}{\partial
v^2}f=J[\mathbf{v}|f,f].
\end{equation}
Here,  $\gamma_\text{b}$ is a drag parameter with a characteristic interaction time $\tau_\text{b,1}^{-1}=\gamma_\text{b}/m$ ($m$ being the mass of a particle) and $\xi_\text{b}^2$ represents the strength of the correlation in the Gaussian white noise of the stochastic force, this having a characteristic interaction time $\tau_\text{b,2}^{-1}=\xi_\text{b}^2/v_0^2$, with $v_0^2=(2T/m)$ and $T$ is the granular temperature. Moreover, the Boltzmann collision operator $J[f,f]$ for IMM is \cite{GS11}
\begin{equation}
\label{2.2} J\left[{\bf v}_{1}|f,f\right] =\frac{\nu(\mathbf{r},t)}{n(\mathbf{r},t)\Omega_d} \int \; \dd{\bf v}_{2}\int
\dd\widehat{\boldsymbol{\sigma}} \left[ \alpha^{-1}f(\mathbf{r},{\bf v}_{1}',t)f(\mathbf{r},{\bf v}_{2}',t)-
f(\mathbf{r}, {\bf v}_{1}, t)f(\mathbf{r}, {\bf v}_{2},t)\right] \;,
\end{equation}
where
\begin{equation}
\label{2.2.1} n({\bf r}, t)=\int \; \dd{\bf v}f({\bf r}, {\bf v},t)
\end{equation}
is the number density, $\Omega_d=2\pi^{d/2}/\Gamma(d/2)$
is the total solid angle in $d$ dimensions and
$\widehat{\boldsymbol{\sigma}}$ is a unit vector along the line of
the two colliding spheres. In addition, the
primes on the velocities denote the initial values $\{{\bf
v}_{1}^{\prime}, {\bf v}_{2}^{\prime}\}$ that lead to $\{{\bf
v}_{1},{\bf v}_{2}\}$ following a binary collision:
\begin{equation}
\label{2.3} {\bf v}_{1}^{\prime}={\bf v}_{1}-\frac{1}{2}\left(
1+\alpha ^{-1}\right)(\widehat{\boldsymbol{\sigma}}\cdot {\bf
g}_{12})\widehat{\boldsymbol {\sigma}}, \quad {\bf v}_{2}^{\prime}={\bf
v}_{2}+\frac{1}{2}\left( 1+\alpha^{-1}\right)
(\widehat{\boldsymbol{\sigma}}\cdot {\bf
g}_{12})\widehat{\boldsymbol{\sigma}}\;,
\end{equation}
where ${\bf g}_{12}={\bf v}_1-{\bf v}_2$ is the relative velocity of the colliding pair.

The collision frequency $\nu(\mathbf{r},t)$ is independent of velocity but depends on space and time through its
dependence on density and temperature. It can be seen as a free parameter of the model that can be chosen to optimize the agreement with the properties of interest of the original Boltzmann equation for IHS. For instance, in order to correctly describe the velocity dependence of the original IHS collision rate, one usually assumes that the IMM collision rate is proportional to $T^q$ with $q=\frac{1}{2}$. Here, the granular temperature is defined as
\begin{equation}
\label{2.6} T({\bf r}, t)=\frac{m}{d n({\bf r}, t)}\int \; \dd{\bf
v} V^2({\bf r}, t) f({\bf r},{\bf v},t),
\end{equation}
where ${\bf V}({\bf r},t)\equiv {\bf v}-{\bf U}({\bf r}, t)$ is the peculiar velocity and
\begin{equation}
\label{2.5}
{\bf U}({\bf r}, t)=\frac{1}{n({\bf r}, t)}\int \;
\dd{\bf v} {\bf v} f({\bf r},{\bf v},t)
\end{equation}
is the mean flow velocity. In this paper, we take $q$ as a generalized exponent so that different values of $q$ can be used to mimic different potentials. As in previous works on IMM \cite{G07,SG07,GT10}, we will assume that $\nu \propto n T^q$, with $q \geq 0$. The case $q=0$ is closer to the original Maxwell model of elastic particles while the case $q=\frac{1}{2}$ is closer to hard spheres. Thus, the collision frequency can be written as \cite{GS11}
\begin{equation}
\label{2.4} \nu=A n T^q,
\end{equation}
where the value of the quantity $A$ will be chosen later.

The macroscopic balance equations for density, momentum, and energy follow directly from Eq.\ ({\ref{2.1}) by multiplying with $1$,
$m{\bf v}$, and $\frac{1}{2}mv^2$ and integrating over ${\bf v}$. The result is
\begin{equation}
\label{2.7} D_{t}n+n\nabla \cdot {\bf U}=0\;,
\end{equation}
\begin{equation}
\label{2.8} D_{t}U_i+(mn)^{-1}\nabla_j P_{ij}=0\;,
\end{equation}
\begin{equation}
\label{2.9} D_{t}T+\frac{2}{dn}\left(\nabla \cdot {\bf
q}+P_{ij}\nabla_j U_i\right) =-\frac{2 T}{m}\gamma_\text{b}+m \xi_\text{b}^2 -\zeta T\;.
\end{equation}
Here, $D_{t}=\partial _{t}+{\bf U}\cdot \nabla$ and the microscopic
expressions for the pressure tensor ${\sf P}$, the heat flux ${\bf
q}$, and the cooling rate $\zeta$ are given, respectively, by
\begin{equation}
{\sf P}({\bf r}, t)=\int \dd{\bf v}\,m{\bf V}{\bf V}\,f({\bf r},{\bf
v},t),
\label{2.10}
\end{equation}
\begin{equation}
{\bf q}({\bf r}, t)=\int \dd{\bf v}\,\frac{1}{2}m V^{2}{\bf V}\,
f({\bf r},{\bf v},t), \label{2.11}
\end{equation}
\begin{equation}
\label{2.12} \zeta({\bf r}, t)=-\frac{1}{dn({\bf r},t)T({\bf r},
t)}\int\, \dd{\bf v} \; m\; V^2\; J[{\bf r},{\bf v}|f(t)].
\end{equation}
The balance equations (\ref{2.7})--(\ref{2.9}) apply regardless of the details of the interaction model considered. The influence of the collision model appears through the $\alpha$-dependence of the cooling rate and of the momentum and heat fluxes.

As said in the Introduction, one of the advantages of the Boltzmann equation for Maxwell models (both elastic and inelastic) is that the collisional moments of the operator $J[f,f]$ can be \emph{exactly} evaluated in terms of the moments of the distribution $f$, without the explicit knowledge of the latter \cite{TM80}. More explicitly, the collisional moments of order $k$ are given as a bilinear combination of moments of order $k'$ and $k''$ with $0\leq k'+k''\leq k$. In particular, the collisional moments involved in the calculation of the momentum and heat fluxes as well as in the fourth cumulant are given by \cite{S03,GS07}
\begin{equation}
\label{2.13}
\int\; \dd\mathbf{v}\; m\; V_iV_j\; J[f,f]=-\nu_{0|2}\left(P_{ij}-p\delta_{ij}\right)-\nu_{2|0} p \delta_{ij},
\end{equation}
\begin{equation}
\label{2.14}
\int\; \dd\mathbf{v}\; \frac{m}{2}\;V^2\;\mathbf{V}\, J[f,f]=-\nu_{2|1}\mathbf{q},
\end{equation}
\begin{equation}
\label{2.15}
\int\; \dd\mathbf{v}\; \;V^4\; J[f,f]=-\nu_{4|0}\langle V^4 \rangle+\lambda_1 d^2\frac{pT}{m^2}-
\frac{\lambda_2}{nm^{2}}\left(P_{ij}-p\delta_{ij}\right)\left(P_{ji}-p\delta_{ij}\right),
\end{equation}
where $p=nT$ is the hydrostatic pressure,
\beq
\nuzt=\frac{(1+\al)(d+1-\al)}{d(d+2)}\nu, \quad \nu_{2|0}=\frac{1-\alpha^2}{2d}\nu,
\label{2.17}
\eeq
\beq
\nuto=\frac{(1+\al)\left[5d+4-\al(d+8)\right]}{4d(d+2)}\nu,
\label{2.18}
\eeq
\beq
\nufz=\frac{(1+\al)\left[12d+9-\alpha(4d+17)+3\alpha^2-3\alpha^3\right]}{8d(d+2)}\nu,
\label{2.19}
\eeq
\beq
\lambda_1=\frac{(1+\al)^2\left(4d-1-6\al+3\al^2\right)}{8d^2}\nu,
\label{2.20}
\eeq
\beq
\lambda_2=\frac{(1+\al)^2\left(1+6\al-3\al^2\right)}{4d(d+2)}\nu,
\label{2.21}
\eeq
and we have introduced the fourth-degree isotropic velocity moment
\beq
\label{2.21}
\langle V^4 \rangle=\int\; \dd \mathbf{v}\; V^4\; f(\mathbf{v}).
\eeq

The cooling rate $\zeta$ for IMM can be determined by taking the trace in Eq.\ \eqref{2.13}. It is given by \cite{S03}
\begin{equation}
\label{2.12.1} \zeta=\frac{1-\alpha^2}{2d}\nu.
\end{equation}
Note that while in the case of IHS, the cooling rate $\zeta$ is also expressed as a functional of the hydrodynamic fields, $\zeta$ is just proportional to $\nu$ in the case of IMM.

In order to compare the results derived here for IMM with those obtained \cite{GCV13} for IHS, we now need a criterion to fix the parameter $\nu$ (or the quantity $A$ in Eq.\ \eqref{2.4}). Analogously to previous works on IMM \cite{S03,G03,GA05,G07,GS07,GS11}, an appropriate choice to optimize the agreement with the IHS results seems to pick $\nu$ as given by Eq.\ \eqref{2.12.1} with $q=\frac{1}{2}$. With this choice, the cooling rate of IMM will be the same as the one obtained for IHS (as evaluated in the Maxwellian approximation) \cite{GS95,NE98}. With this choice, the collision frequency $\nu$ is
\begin{equation}
\label{2.12.2}
\nu=\frac{d+2}{2}\nu_0,
\end{equation}
where
\begin{equation}
\label{2.12.3}
\nu_0=\frac{4\Omega_d}{\sqrt{\pi}(d+2)}n \sigma^{d-1} \sqrt{\frac{T}{m}}.
\end{equation}
The collision frequency $\nu_0$ is the one associated with the Navier-Stokes shear viscosity of an ordinary gas ($\al=1$) of both Maxwell molecules and hard spheres, i.e., $\eta_0=p/\nu_0$.

\section{Homogeneous steady states}
\label{sec3}

Before analyzing inhomogeneous states, it is quite convenient first to study the homogeneous problem. In this case, the density $n$ is constant, the flow velocity vanishes and the temperature $T(t)$ is spatially uniform. Consequently, the Boltzmann equation \eqref{2.1} becomes
\begin{equation}
\partial_{t}f-\frac{\gamma_\text{b}}{m}
\frac{\partial}{\partial{\bf v}}\cdot {\bf v}
f-\frac{1}{2}\xi_\text{b}^2\frac{\partial^2}{\partial v^2}f=J[f,f]. \label{3.1}
\end{equation}
Since the heat flux vanishes and the pressure tensor is diagonal ($P_{ij}=p\delta_{ij}$), then the energy balance equation \eqref{2.9} reads simply
\begin{equation}
\label{3.2}
\partial_tT=-\frac{2 T}{m}\gamma_\text{b} +m \xi_\text{b}^2-\zeta \,T.
\end{equation}
In the hydrodynamic regime, the time dependence of $f$ only occurs through the relevant fields. In the homogeneous state, the only (time) varying field is the granular temperature $T$:
\begin{equation}
\label{3.3}
\partial_t f=\frac{\partial
f}{\partial T}\partial_tT= -\left(\frac{2}{m}\gamma_\text{b} -\frac{m}{T} \xi_\text{b}^2+\zeta\right) T\frac{\partial f}{\partial T}.
\end{equation}
Substitution of Eq.\ (\ref{3.3}) into Eq.\ (\ref{3.1}) yields
\beq
\label{3.4}
-\left( \frac{2}{m}\gamma_\text{b}-\frac{m}{T}\xi_\text{b}^2 + \zeta\right)T
\frac{\partial f}{\partial T}-\frac{\gamma_\text{b}}{m} \frac{\partial}{\partial {\bf
v}}\cdot {\bf v} f-\frac{1}{2}\xi_\text{b}^2\frac{\partial^2}{\partial
v^2}f=J[f,f].
\eeq

For ordinary (elastic) gases ($\al=1$), $\zeta=0$ and the solution to Eq.\ \eqref{3.4} is the Maxwellian distribution
\beq
\label{max}
f_\text{M}(v)=n\left(\frac{m}{2\pi T_\text{b}}\right)^{d/2}\; \exp\left(-\frac{mv^2}{2T_\text{b}}\right)
\eeq
where
\beq
\label{3.5.1}
T_\text{b}=\frac{m^2\xi_\text{b}^2}{2\gamma_\text{b}}
\eeq
is the temperature of the (equilibrium) background bath \cite{GSVP11}. The relation \eqref{3.5.1} is a consequence of the well-known fluctuation-dissipation theorem \cite{K92} relating the dissipation resulting from the action of an external force to the spontaneous fluctuations at thermal equilibrium. For granular gases ($\alpha \neq 1$, and so $\zeta \neq 0$), the fluctuation-dissipation theorem does not strictly apply and hence, the bath is not at equilibrium. In this case, the drag coefficient $\gamma_\text{b}$ and the amplitude of the stochastic force $\xi_\text{b}^2$ are generally not related (namely, they can be chosen as independent parameters). On the other hand, as we will show later, we shall consider a relation between both parameters (see Eq.\ \eqref{4.10} below) to simplify the calculations performed to determine the transport coefficients.

In the \emph{steady} state, the first term on the left hand side of Eq.\ \eqref{3.4} vanishes and the steady temperature $T_\text{s}$ is given by
\begin{equation}
\label{3.5}
\zeta_\text{s} T_\text{s}+ \frac{2\gamma_\text{b}}{m} T_\text{s} =m\xi_\text{b}^2,
\end{equation}
where subscript $\text{s}$ stands for the steady state. By combining relations \eqref{3.4} and \eqref{3.5} we can write, for the steady state,
\begin{equation}
\label{3.6}
\frac{1}{2}\zeta_\text{s} \frac{\partial}{\partial {\bf
v}}\cdot {\bf v} f_\text{s}-\frac{m\xi_\text{b}^2}{2T_\text{s}} \frac{\partial}{\partial {\bf
v}}\cdot {\bf v} f_\text{s}-\frac{1}{2}\xi_\text{b}^2\frac{\partial^2}{\partial v^2}f_\text{s}=J[f_\text{s},f_\text{s}].
\end{equation}

Equation \eqref{3.6} shows that $f_\text{s}$ depends on the driven parameter $\xi_\text{b}^2$. Thus, dimensionless analysis requires that $f_\text{s}$ has the form \cite{CVG13}
\begin{equation}
\label{3.7}
f_\text{s}({\bf v}, \xi_\text{b}^2)=n_\text{s}v_0^{-d}\varphi_\text{s}\left(\mathbf{c}, \xi_\text{s}^*\right),
\end{equation}
where $\varphi_\text{s}$ is an unknown function of the dimensionless  parameters
\begin{equation}
\label{3.8}
\mathbf{c}\equiv \frac{\mathbf {v}}{v_0}, \quad \xi_\text{s}^*=\frac{m\xi_\text{b}^2}{T_\text{s}\nu_\text{s}},
\end{equation}
where $v_0=\sqrt{2T_\text{s}/m}$ is the thermal velocity and $\nu_\text{s}=A n_\text{s} T_\text{s}^q$.

As already noted in previous studies of IHS \cite{CVG13,GMT12}, the scaled distribution $\varphi_\text{s}$ depends on the granular temperature through the scaled velocity $\mathbf{c}$ and \emph{also} through the (reduced) noise strength $\xi_s^*$. On the contrary, in the homogeneous cooling state and in the case of only one thermostat force, the dependence of $\varphi_\text{s}$ is only encoded by the single parameter $\mathbf{c}$ \cite{S03}. In dimensionless form, Eq.\ \eqref{3.6} can be written as
\begin{equation}
\label{3.9}
\frac{1}{2}(\zeta_s^* - \xi_s^*) \frac{\partial}{\partial\mathbf{c}} \cdot \mathbf{c} \varphi_\text{s}- \frac{1}{4}\xi_s^* \frac{\partial^2}{\partial c^2} \varphi_\text{s} = J^*[\varphi_\text{s},\varphi_\text{s}],
\end{equation}
where $\zeta_s^*\equiv \zeta_s/\nu_\text{s}=(1-\al^2)/2d$ and $J^*[\varphi_\text{s},\varphi_\text{s}]\equiv v_0^d J[f_\text{s},f_\text{s}]/(n_\text{s} \nu_\text{s})$.

In reduced units, the steady state condition \eqref{3.5} can be written as
\beq
\label{3.9.1}
2\gamma_\text{s}^*=\xi_\text{s}^*-\zeta_\text{s}^*,
\eeq
where $\gamma_\text{s}^*\equiv \gamma_\text{b}/(m \nu_\text{s})$. Since $\gamma_\text{s}^*$ is definite positive, then Eq.\ \eqref{3.9.1} requires that $\xi_\text{s}^*\geq \zeta_\text{s}^*$. Thus, at a given value of $\al$, there is a minimum threshold value $\xi_\text{th}^*(\al)=\zeta_s^*$ needed to achieve a steady state. In particular, for spheres ($d=3$), the smallest value of $\xi_\text{th}^*(\al)$ is $1/6$ (which corresponds to $\al=0$) while the smallest value of $\xi_\text{th}^*(\al)$ for disks ($d=2$) is $1/4$.
\begin{figure}
{\includegraphics[width=0.47\columnwidth]{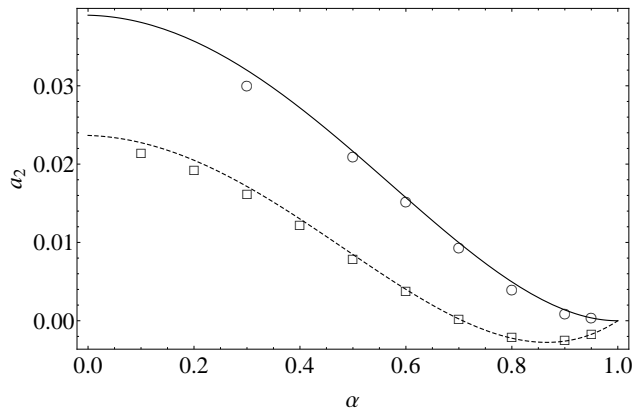}}
\caption{The (steady) fourth-cumulant $a_{2,\text{s}}$ as a function of the coefficient of restitution for a thre-dimensional system ($d=3$) for $\xi_\text{s}^*=0.62$. The solid and dashed lines are the analytic results obtained for IMM and IHS, respectively. The symbols refer to the Monte Carlo simulation results for IMM (circles) and IHS (squares).  \label{fig1}}
\end{figure}

In the case of elastic collisions ($\al=1$), $\zeta_\text{s}^*=0$ and the solution to Eq.\ \eqref{3.9} is the gaussian distribution $\varphi_\text{M}(c)=\pi^{-d/2}e^{-c^2}$. On the other hand, if $\al\neq 1$, then $\zeta_\text{s}^*\neq 0$ and the solution to Eq.\ \eqref{3.9} is not exactly known. An indirect information of the deviation of $\varphi_\text{s}(c)$ from its gaussian form $\varphi_\text{M}(c)$ is given by the kurtosis or fourth-cumulant
\begin{equation}
\label{3.10}
a_{2,\text{s}}=\frac{4}{d(d+2)}\langle c^4\rangle-1,
\end{equation}
where
\begin{equation}
\label{3.11}
\langle c^k\rangle=\int\; \dd{\bf c}\; c^k \varphi_s(c).
\end{equation}
In order to determine $a_{2,\text{s}}$, we multiply Eq.\ \eqref{3.9} by $c^4$ and integrate over velocity. The result is
\begin{equation}
\label{3.12}
2(\zeta_s^*-\xi_s^*)\left(
1+a_{2,\text{s}}\right)+2\xi_s^*=\left(
1+a_{2,\text{s}}\right)\nu_{4|0}^*-\frac{d}{(d+2)}\lambda_1^*,
\end{equation}
where $\nu_{4|0}^*\equiv \nu_{4|0}/\nu_\text{s}$, $\lambda_1^*\equiv \lambda_1/\nu_\text{s}$ and use has been made of Eq.\ \eqref{2.15}. The solution to Eq.\ \eqref{3.12} is
\beq
\label{3.13}
a_{2,\text{s}}=\frac{2\zeta_s^*-\nu_{4|0}^*+\frac{d}{d+2}\lambda_1^*}
{\nu_{4|0}^*-2(\zeta_s^*-\xi_s^*)}=\frac{6(1-\al^2)^2}{4d-7+3\al(2-\al)+16d(d+2)\xi_\text{s}^*},
\eeq
where the explicit forms of $\zeta_s^*$ and $\nu_{4|0}^*$ have been considered. In the absence of friction ($\gamma_\text{b}=0$), the steady state condition \eqref{3.9.1} becomes $\xi_\text{s}^*=\zeta_s^*$ and we have double checked that Eq.\ \eqref{3.13} yields back the results of the theory of a driven granular gas heated only by the stochastic thermostat \cite{S03}
\beq
\label{3.14}
a_{2,\text{s}}=\frac{6(1-\al)^2(1+\al)}{12d+9-\alpha(4d+17)+3\alpha^2(1-\alpha)}.
\eeq
Moreover, when $\xi_\text{s}^*=0$, Eq.\ \eqref{3.13} is consistent with the one obtained for IMM in the freely cooling case \cite{S03}.

Figure \ref{fig1} shows the steady value of the fourth-cumulant $a_{2,\text{s}}$ versus the coefficient of restitution $\al$ for a three-dimensional system. The theoretical results derived here for IMM given by Eq.\ \eqref{3.13} and in Ref.\ \cite{CVG13} (see Eq.\ \eqref{b7}) for IHS are compared with those obtained by numerically solving the Boltzmann equation for IMM and IHS, respectively, by means of the direct simulation Monte Carlo (DSMC) method \cite{B94}. The parameters of the simulations for IMM and IHS have been chosen to get $\xi_\text{s}^*=0.62$ in the steady state. It is seen that the homogeneous state of IHS deviates from the gaussian distribution $\varphi_\text{M}(c)$ (which corresponds to $a_2=0$) more than the homogeneous state of IHS. This behavior contrasts with the results obtained in the freely cooling case \cite{S03} where the magnitude of $a_2$ for IMM is much larger than that of IHS. As expected, the simulation data for IMM and IHS show an excellent agrement with the exact result for IMM  (Eq.\ \eqref{3.13}) and with the fist Sonine approximation for IHS (Eq.\ \eqref{b7}), even for quite small values of $\al$.

\section{Chapman-Enskog method for states close to homogeneous steady states}
\label{sec4}

Let us slightly disturb the homogeneous steady state by small spatial perturbations. In this case, the momentum and heat fluxes are not zero and their corresponding Navier-Stokes transport coefficients can be identified. The evaluation of these coefficients as functions of the coefficient of restitution and the parameters of the external force is the main goal of the present paper.

As long as the spatial gradients keep small, the Boltzmann equation \eqref{2.1} may be solved by means of the Chapman-Enskog method \cite{CC70} adapted to inelastic collisions. The Chapman-Enskog method assumes the existence of a \emph{normal} solution in which all the space and time dependence of the distribution function occurs only through a functional dependence on the hydrodynamic fields, i.e.,
\begin{equation}
f({\bf r},{\bf v},t)=f\left[{\bf v}|n ({\bf r}, t),
T({\bf r}, t), {\bf U}({\bf r}, t) \right] \;.
\label{4.1}
\end{equation}
The notation on the right hand side indicates a functional dependence on the density, temperature and flow velocity. This functional dependence can be made local by an expansion of $f({\bf r},{\bf v},t)$ in powers of the spatial gradients of $n$, $\mathbf{U}$, and $T$:
\begin{equation}
f=f^{(0)}+f^{(1)}+f^{(2)}+\cdots \;, \label{4.2}
\end{equation}
where the approximation $f^{(k)}$ is of order $k$ in spatial gradients. In addition, to collect the different level of approximations in Eq.\ \eqref{2.1}, one has to characterize the magnitude of the external driven parameters with respect to the gradients as well. As in Ref.\ \cite{GCV13}, we assume that the parameters $\gamma_\text{b}$ and $\xi_\text{b}^2$ are taken to be of zeroth order in gradients since they do not create any new contribution to the irreversible fluxes and only modify the form of the transport coefficients.

The expansion \eqref{4.2} yields the corresponding expansions for the fluxes when one substitutes \eqref{4.2} into their definitions \eqref{2.10} and \eqref{2.11}:
\beq
\label{4.3}
{\sf P}={\sf P}^{(0)}+{\sf P}^{(1)}+\ldots, \quad \mathbf{q}=\mathbf{q}^{(0)}+\mathbf{q}^{(1)}+\ldots.
\eeq
Note that the cooling rate is exactly given by the expression \eqref{2.12.1} and so, $\zeta^{(k)}=0$ for $k\geq 0$. In the case of IHS, $\zeta^{(1)}$ is different from zero but very small \cite{GCV13}. Finally, as usual in the Chapman-Enskog method, the time derivative is also expanded as
\beq
\label{4.4}
\partial_t=\partial_t^{(0)}+\partial_t^{(1)}+\ldots,
\eeq
where the action of each operator $\partial_t^{(k)}$ is obtained from the macroscopic balance equations \eqref{2.7}--\eqref{2.9} when one represents the fluxes and the cooling rate in their corresponding series expansion \eqref{4.3}. In this paper, only the first order contributions to the fluxes will be considered.

\subsection{Zeroth-order approximation}

Substitution of Eqs.\ (\ref{4.2})--(\ref{4.4}) into Eq.\ (\ref{2.1}), yields the kinetic equation for $f^{(0)}$
\begin{equation}
\label{4.5}
\partial_{t}^{(0)}f^{(0)}-\frac{\gamma_\text{b}}{m}
\frac{\partial}{\partial{\bf v}}\cdot {\bf V}
f^{(0)}-\frac{1}{2}\xi_\text{b}^2\frac{\partial^2}{\partial v^2}f^{(0)}=J[{\bf V}|f^{(0)},f^{(0}].
\end{equation}
To lowest order in the expansion, the balance equations yield
\begin{equation}
\label{4.8}
\partial_t^{(0)}n=0,\quad \partial_t^{(0)}\mathbf{U}=\mathbf{0}, \quad \partial_t^{(0)}T=-\frac{2T}{m}\gamma_\text{b} +m\xi_\text{b}^2-
\zeta T,
\end{equation}
where $\zeta$ is given by Eq.\ \eqref{2.12.1}. Note that the cooling rate depends on space and time through the density $n(\mathbf{r},t)$ and temperature $T(\mathbf{r},t)$ fields. Moreover, $\partial_t^{(0)}f^{(0)} \to
(\partial_T f^{(0)})(\partial_t^{(0)} T)$ and thus, Eq. \eqref{4.5} becomes
\beq
-\left(\frac{2}{m}\gamma_\text{b} -\frac{m}{T} \xi_\text{b}^2+\zeta\right) T\frac{\partial f^{(0)}}{\partial T}-\frac{\gamma_\text{b}}{m}
\frac{\partial}{\partial{\bf v}}\cdot {\bf V}
f^{(0)}-\frac{1}{2}\xi_\text{b}^2\frac{\partial^2}{\partial v^2}f^{(0)}=J[f^{(0)},f^{(0)}].
\label{4.9}
\eeq
As already noted in the case of IHS \cite{GCV13}, since density and temperature are specified separately in the \emph{local} reference state $f^{(0)}$, the collisional cooling and the action of the thermostats do not in general cancel each other at all points in the system. Thus, $\partial_t^{(0)} T \neq 0$ and $f^{(0)}$ depends on time through its dependence on the temperature.

As said before, in the case of elastic collisions, the fluctuation-dissipation theorem yields Eq.\ \eqref{3.5.1} where $T_\text{b}$ is the bath temperature. In the case of inelastic collisions, the above theorem does not hold and the model parameters $\gamma_\text{b}$ and $\xi_\text{b}^2$ does not necessarily obey the relation \eqref{3.5.1}. However, to simplify the calculations in the time-dependent problem, we assume that those parameters verify the generic relation
\begin{equation}
\label{4.10}
\gamma_\text{b}=\beta \frac{m^2\xi_\text{b}^2} {T_\text{b}},
\end{equation}
where $\beta$ is a constant and $T_\text{b}$ is an arbitrary (known) temperature. Here, to make contact with some works \cite{GSVP11} that have previously used the kind of thermostat considered in this paper, we have taken $T_\text{b}$ as the temperature of the background bath when the latter is at equilibrium. The relation \eqref{4.10} was also assumed in the previous work for IHS \cite{GCV13}.
When $\beta=0$ (or equivalently, when $\gamma_\text{b}=0$ but $\gamma_\text{b}T_\text{b}\equiv \text{finite}$) our thermostat reduces to the usual stochastic thermostat \cite{WM96,GMT12} while the choice $\beta=\frac{1}{2}$ yields back the conventional Fokker-Planck model \cite{GSVP11,K92,RL,H03}. Thus, we will consider henceforth these two physically relevant values ($\beta=0,1/2$). Equation \eqref{4.10} can be rewritten as
\begin{equation}
\label{4.11}
\gamma^*=\beta T^* \xi^*=\theta \xi^{*q/(1+q)},
\end{equation}
where $T^*\equiv T/T_\text{b}$ and
\begin{equation}
\label{4.12}
\theta\equiv \beta \left(\frac{m \xi_\text{b}^2}{A n T_\text{b}^{1+q}}\right)^{1/(1+q)}.
\end{equation}
Upon writing Eq.\ \eqref{4.11}, use has been made of the identity $\beta T^*=\theta/\xi^{*1/(1+q)}$, where
\begin{equation}
\label{4.12.1}
\xi^*\equiv \frac{m\xi_\text{b}^2}{T \nu(T)}=\frac{m\xi_\text{b}^2}{A n T^{q+1}}.
\end{equation}
According to Eq.\ \eqref{4.12}, for the simplest model $q=0$, $\theta$ can be interpreted as the (dimensionless) white noise intensity reduced with respect to the \emph{bath} temperature $T_\text{b}$. On the other hand, the (dimensionless) noise strength $\xi^*$ has been reduced with respect to the actual \emph{granular} temperature $T$. Note that $\theta$ depends on space through its dependence on $n$ while $\xi^*$ depends on space through its dependence on $n$ and $T$.

In the \emph{unsteady} state, dimensional analysis also requires that the zeroth-order distribution $f^{(0)}(\mathbf{r}, \mathbf{v}, t)$ has the scaled form \eqref{3.7} (once one uses the relation \eqref{4.11}), namely
\beq
\label{4.13}
f^{(0)}({\bf r}, {\bf v}, t)=n({\bf r},t) v_0({\bf r},t)^{-d}\varphi\left(\mathbf{c}, \theta, \xi^*\right),
\eeq
where now $\mathbf{c}\equiv \mathbf{V}/v_0$, $\mathbf{V}={\bf v}-{\bf U}$ being the peculiar velocity. The temperature dependence of the reduced distribution  $\varphi$ is encoded by the dimensionless velocity $\mathbf{c}$ and the (reduced) noise strength $\xi^*$. Consequently, according to Eq.\ \eqref{4.13}, one gets
\begin{equation}
\label{4.14}
T\frac{\partial}{\partial T} f^{(0)} =-\frac{1}{2}\frac{\partial}{\partial \mathbf{V}} \cdot \mathbf{V} f^{(0)} -(1+q) \xi^* \frac{\partial}{\partial \xi^*} f^{(0)},
\end{equation}
and the scaled distribution $\varphi$ obeys the kinetic equation
\begin{equation}
\label{4.15} (1+q)\left[ \left( 2\beta T^* -1 \right) \xi^* + \zeta^* \right] \xi^*
\frac{\partial \varphi}{\partial \xi^*}+
\frac{1}{2}(\zeta^* - \xi^*) \frac{\partial}{\partial\mathbf{c}} \cdot \mathbf{c}\varphi
- \frac{1}{4}\xi^* \frac{\partial^2 \varphi}{\partial c^2}
= J^*[\varphi,\varphi],
\end{equation}
where use has been made of the identity \eqref{4.11}.

An implicit expression of the fourth-cumulant $a_2(\xi^*)$ (defined by Eq.\ \eqref{3.10}) can be obtained for unsteady states by multiplying both sides of Eq.\ \eqref{4.15} by $c^4$ and integrating over velocity. The result is
\beq
\label{4.16}
(1+q)\left[ \left( 2\beta T^* -1 \right) \xi^* +\zeta^*\right]\xi^*
\frac{\partial a_2}{\partial\xi^*}=\frac{d}{d+2}\lambda_1^*+(1+a_2)(2\zeta^*-\nu_{4|0}^*) -2\xi^* a_2.
\eeq
In Eq.\ \eqref{4.16}, the function $a_2(\xi^*)$ must be obtained numerically. As we will show later, evaluation of the Navier-Stokes transport coefficients in the steady state requires the knowledge of the derivatives
$\partial a_2/\partial \xi^*$ and $\partial a_2/\partial \theta$ in this state.

\subsection{First-order approximation}

The analysis to first order in the gradients follows similar steps as those made for IHS \cite{GCV13}.  The velocity distribution function $f^{(1)}$ verifies the kinetic equation
\beq
\label{4.21}
\left(\partial_{t}^{(0)}+{\cal L}\right)f^{(1)}-\frac{\gamma_\text{b}}{m}
\frac{\partial}{\partial {\bf v}}\cdot {\bf V}f^{(1)}-\frac{1}{2}\xi_\text{b}^2\frac{\partial^2}{\partial v^2}f^{(1)}=-\left(\partial_{t}^{(1)}+{\bf v}\cdot \nabla \right)f^{(0)},
\eeq
where  ${\cal L}$ is the linearized Boltzmann collision operator
\begin{equation}
\label{4.22}
{\cal L}f^{(1)}=-\left(J[f^{(0)},f^{(1)}]+J[f^{(1)},f^{(0)}]\right).
\end{equation}
The macroscopic balance equations \eqref{2.7}--\eqref{2.9} to first order in the gradients are
\begin{equation}
\label{4.23}
D_t^{(1)}n=-n\nabla\cdot {\bf U},\quad
D_t^{(1)}U_i=-(mn)^{-1}\nabla_i p,
\end{equation}
\begin{equation}
\label{4.24}
D_t^{(1)}T=-\frac{2p}{dn}\nabla\cdot {\bf U},
\end{equation}
where $D_t^{(1)}\equiv \partial_t^{(1)}+{\bf U}\cdot \nabla$ and $p=nT$ is the hydrostatic pressure. Use of Eqs.\ (\ref{4.23}) in Eq.\ (\ref{4.21}) leads to
\beqa
\label{4.25}
 \left(\partial_{t}^{(0)}+{\cal L}\right)f^{(1)}&-&\frac{\gamma_\text{b}}{m}
\frac{\partial}{\partial {\bf v}}\cdot {\bf V}
f^{(1)}-\frac{1}{2}\xi_\text{b}^2\frac{\partial^2}{\partial v^2}f^{(1)}
={\bf A}\cdot
\nabla \ln T+{\bf B}\cdot \nabla \ln n\nonumber\\
&+&C_{ij}\frac{1}{2}\left( \nabla_{i}U_{j}+\nabla_{j}U_{i}-\frac{2}{d}\delta _{ij}\nabla \cdot\mathbf{U} \right)+D \nabla \cdot\mathbf{U},
\eeqa
where
\begin{equation}
{\bf A}\left( \mathbf{V}\right)=-\mathbf{V}T\frac{\partial f^{(0)}}{\partial T}
-\frac{p}{\rho }\frac{\partial f^{(0)}}{\partial \mathbf{V}},  \label{4.26}
\end{equation}
\beq
{\bf B}\left(\mathbf{V}\right)= -{\bf V}n\frac{\partial f^{(0)}}{\partial n}-\frac{p}{\rho}
\frac{\partial f^{(0)}}{\partial \mathbf{V}},  \label{4.27}
\eeq
\begin{equation}
\label{4.28}
C_{ij}\left(
\mathbf{V}\right)=V_i\frac{\partial f^{(0)}}{\partial V_j},
\end{equation}
\beq
D=\frac{1}{d}\frac{\partial}{\partial \mathbf{V}}\cdot (\mathbf{V}
f^{(0)})+\frac{2}{d}T\frac{\partial f^{(0)}}{\partial T}-f^{(0)}+n\frac{\partial f^{(0)}}{\partial n}.   \label{4.29}
\eeq
In Eqs.\ \eqref{4.24} and \eqref{4.25}, $T\partial_T f^{(0)}$ is given by Eq.\ \eqref{4.14} while, according to Eqs.\ \eqref{4.10} and \eqref{4.11}, the term $n\partial_n f^{(0)}$ can be more explicitly written as
\begin{equation}
\label{4.29.1}
n\frac{\partial f^{(0)}}{\partial n}=f^{(0)}-\xi^*\frac{\partial f^{(0)}}{\partial \xi^*}
-\frac{\theta}{1+q}\frac{\partial f^{(0)}}{\partial \theta}.
\end{equation}
It is worth noticing that for $q=\frac{1}{2}$, Eqs.\ \eqref{4.25}--\eqref{4.29} have the same structure as that of the Boltzmann equation for IHS \cite{GCV13}. The only difference between both models lies in the explicit form of the linearized operator ${\cal L}$.

\section{Navier-Stokes transport coefficients}
\label{sec5}

This section is devoted to the calculation of the Navier-Stokes transport coefficients of the driven granular gas. These coefficients can be identified from the expressions of the first-order contributions to the pressure tensor
\beq
\label{5.1}
{\sf P}^{(1)}=\int\; \dd \mathbf{v}\; m \mathbf{V} \mathbf{V} f^{(1)}(\mathbf{V}),
\eeq
and the heat flux vector
\beq
\label{5.2}
\mathbf{q}^{(1)}=\int\; \dd \mathbf{v}\; \frac{m}{2} V^2 \mathbf{V} f^{(1)}(\mathbf{V}).
\eeq

The evaluation of the above fluxes has been worked out in Appendix \ref{appA}. Only the final results are presented in this section. The pressure tensor $P_{ij}^{(1)}$ is given by
\begin{equation}
\label{5.2.1}
P_{ij}^{(1)}=-\eta\left( \nabla_{i}U_{j}+\nabla_{j
}U_{i}-\frac{2}{d}\delta _{ij}\nabla \cdot
\mathbf{U} \right),
\end{equation}
while the heat flux $\mathbf{q}^{(1)}$ is
\beq
\label{5.2.2}
\mathbf{q}^{(1)}=-\kappa \nabla T-\mu \nabla n.
\eeq
Here, $\eta$ is the shear viscosity coefficient, $\kappa$ is the thermal conductivity coefficient and $\mu$ is a new transport coefficient not present for ordinary gases. These transport coefficients can be written in the form
\beq
\label{5.3}
\eta=\eta_0 \eta^*, \quad \kappa=\kappa_0 \kappa^*, \quad \mu=\frac{\kappa_0 T}{n}\mu^*,
\eeq
where $\eta_0=(d+2)(p/2\nu)$ and $\kappa_0=[d(d+2)/2(d-1)](\eta_0/m)$ are the shear viscosity and thermal conductivity coefficients, respectively, of a dilute ordinary gas. The reduced coefficients $\eta^*$, $\kappa^*$ and $\mu^*$ depend on temperature through its dependence on the (reduced) noise strength $\xi^*$. They verify the following first-order differential equations:
\beq
\label{5.4}
\Lambda^*\left[(1-q)\eta^*-(1+q)\xi^*\frac{\partial \eta^*}{\partial \xi^*}\right]+
\left(\nu_{0|2}^*+2\gamma^*\right)\eta^*=\frac{2}{d+2},
\eeq
\beq
\label{5.5}
\Lambda^*\left[(1-q)\kappa^*-(1+q)\xi^*\frac{\partial \kappa^*}{\partial \xi^*}\right]+\left(\Lambda^*-\xi^*-q \zeta^*+\nu_{2|1}^*+3\gamma^*\right)\kappa^* =\frac{2(d-1)}{d(d+2)}\left[1+2a_2-
(1+q)\xi^*\frac{\partial a_2}{\partial \xi^*}\right],
\eeq
\beq
\label{5.6}
\Lambda^*\left[(2-q)\mu^*-(1+q)\xi^*\frac{\partial \mu^*}{\partial \xi^*}\right]+\left(\nu_{2|1}^*+3\gamma^*\right)\mu^*=\zeta^*\kappa^*
+\frac{2(d-1)}{d(d+2)}\left(a_2-\frac{\theta}{1+q}\frac{\partial a_2}{\partial \theta}-\xi^*\frac{\partial a_2}{\partial \xi^*}\right).
\eeq
Here,
\beq
\label{5.7}
\Lambda^*=\xi^*-2\gamma^*-\zeta^*,
\eeq
$\nu_{0|2}^*\equiv \nu_{0|2}/\nu$ and $\nu_{2|1}^*\equiv \nu_{2|1}/\nu$, where $\nu_{0|2}$ and $\nu_{2|1}$ are given by Eqs.\ \eqref{2.17} and \eqref{2.18}, respectively.

Apart from the Navier-Stokes transport coefficients (which are directly related to the second- and third-degree velocity moments of the the first order distribution function $f^{(1)}$), another interesting velocity moment of $f^{(1)}$ corresponds to its fourth degree isotropic moment defined as
\beq
\label{5.8}
e_D=\frac{1}{2d(d+2)}\frac{m^2}{nT^2}\int\; \dd\mathbf{v}\; V^4 f^{(1)}.
\eeq
In dimensionless form, the coefficient $e_D$ is given by
\beq
\label{5.9}
e_D=e_D^* \nu^{-1} \nabla \cdot \mathbf{U},
\eeq
where $e_D^*$ is the solution of the first-order differential equation
\beq
\label{5.10}
\Lambda^*\left[(2-q)e_D^*-(1+q)\xi^*\frac{\partial e_D^*}{\partial \xi^*}\right]+\left(\nu_{4|0}^*+4\gamma^*\right)e_D^*=-\frac{2(1+q)+d}{2d}\xi^*\frac{\partial a_2}{\partial \xi^*}
-\frac{1}{2}\frac{\theta}{1+q}\frac{\partial a_2}{\partial \theta}.
\eeq
Here, $\nu_{4|0}^*\equiv \nu_{4|0}/\nu$ where $\nu_{4|0}$ is given by Eq.\ \eqref{2.19}.

In the elastic limit ($\al=1$), $\zeta_\text{s}^*=0$, $a_{2,\text{s}}=0$, $\gamma_\text{s}^*=\xi_\text{s}^*/2$, $\nu_{0|2}^*=2/(d+2)$, and $\nu_{2|1}^*=2(d-1)/d(d+2)$. In this case, $\mu_\text{s}^*=e_D^*=0$ and the coefficients $\eta_\text{s}^*$ and $\kappa_\text{s}^*$ become, respectively,
\beq
\label{5.19}
\eta_\text{s}^*\to \eta_\text{s,0}^*=\frac{1}{1+\frac{d+2}{2}\xi_\text{s}^*},\quad
\kappa_\text{0}^*\to \kappa_\text{s,0}^*=\frac{1}{1+\frac{d(d+2)}{4(d-1)}\xi_\text{s}^*}.
\eeq
Another interesting situation is the freely cooling gas ($\gamma^*=\xi^*=0$). In this case, $\Lambda^*=-\zeta^*$ and Eq.\ \eqref{5.10} gives $e_D^*=0$. In addition, the solution to Eqs.\ \eqref{5.4}--\eqref{5.6} can be written as
\beq
\label{5.11}
\eta^*=\frac{2}{d+2}\frac{1}{\nu_{0|2}^*-(1-q)\zeta^*},
\eeq
\beq
\label{5.12}
\kappa^*=\frac{2(d-1)}{d(d+2)}\frac{1+2a_2}{\nu_{2|1}^*-2\zeta^*},
\eeq
\beq
\label{5.13}
\mu^*=\frac{\kappa^*}{1+2a_2}\frac{\zeta^*+\nu_{2|1}^*a_2}{\nu_{2|1}^*-\left(2-q\right)\zeta^*}.
\eeq
When $q=\frac{1}{2}$, Eqs.\ \eqref{5.11}--\eqref{5.13} agree with those previously derived \cite{S03} for an undriven granular gas of IMM.

Apart from the above two situations (elastic collisions and undriven granular gas), the evaluation of the transport coefficients ($\eta^*$, $\kappa^*$, $\mu^*$, and $e_D^*$ ) for the general case of unsteady states requires to solve the differential equations \eqref{5.4}--\eqref{5.6} and \eqref{5.10}. However, even for the simplest model ($q=0$), it is not possible to obtain an exact solution to this system of equations, except in the steady state limit. For the steady state ($\Lambda^*=0$), one has to still evaluate the derivatives of $\partial a_2/\partial\xi^*$ and $\partial a_2/\partial\theta$. The steady state expressions of these derivatives may be easily deduced, as we will show, from the simplified steady state form of Eq. \eqref{4.16}. We present the results for steady states in the next subsection.

\subsection{Transport coefficients under steady state}

Under steady state ($\Lambda^*=0$), the set of differential equations \eqref{5.4}--\eqref{5.6} and \eqref{5.10} become a simple set of algebraic equations whose solution is
\beq
\label{5.14}
\eta_\text{s}^*=\frac{2}{d+2}\frac{1}{\nu_{0|2}^*+2\gamma_\text{s}^*},
\eeq
\beq
\label{5.15}
\kappa_\text{s}^*=\frac{2(d-1)}{d(d+2)}\frac{1+2 a_{2,\text{s}}-(1+q)\xi_\text{s}^* \left(\frac{\partial a_2}{\partial \xi^*}\right)_{\text{s}}}{\nu_{2|1}^*+\frac{1}{2}\xi_\text{s}^*-\left(q+\frac{3}{2}\right)
\zeta_\text{s}^*},
\eeq
\beq
\label{5.16}
\mu_\text{s}^*=\frac{
\zeta_\text{s}^*\kappa_\text{s}^*+\frac{2(d-1)}{d(d+2)}\left[
a_{2,\text{s}}-\frac{\theta_\text{s}}{1+q}\left(\frac{\partial a_2}{\partial \theta}\right)_\text{s}-\xi_\text{s}^*\left(\frac{\partial a_2}{\partial \xi^*}\right)_\text{s}\right]}
{\nu_{2|1}^*+3\gamma_\text{s}^*},
\eeq
\beq
\label{5.17}
e_D^*=-\frac{\frac{2(1+q)+d}{2d}\xi_\text{s}^*\left(\frac{\partial a_2}{\partial \xi^*}\right)_\text{s}
+\frac{1}{2}\frac{\theta_\text{s}}{1+q}\left(\frac{\partial a_2}{\partial \theta}\right)_\text{s}}
{\nu_{4|0}^*+4\gamma_\text{s}^*},
\eeq
where $\gamma_\text{s}^*=(\xi_\text{s}^*-\zeta_\text{s}^*)/2$ and
\beq
\label{5.18}
\theta_\text{s}=\frac{\xi_\text{s}^*-\zeta_\text{s}^*}{2}\xi_\text{s}^{*q/(1+q)}.
\eeq

The derivatives $(\partial a_2/\partial \xi^*)_\text{s}$ and $(\partial a_2/\partial \theta)_\text{s}$ appearing in Eqs.\ \eqref{5.15}--\eqref{5.17} can be easily obtained from Eq. \eqref{4.16}.  According to Eq.\ \eqref{4.16}, the derivative $\partial a_2/\partial \xi^*$ is given by
\beq
\label{4.16.1}
\frac{\partial a_2}{\partial \xi^*}=\frac{\frac{d}{d+2}\lambda_1^*+(1+a_2)(2\zeta^*-\nu_{4|0}^*) -2\xi^* a_2}
{(1+q)\xi^*\left[ \left( 2\beta T^* -1 \right) \xi^* +\zeta^*\right]}.
\eeq
In the steady state, the numerator and denominator of Eq.\ \eqref{4.16.1} vanish so that, the quantity $\partial a_2/\partial \xi^*$ becomes indeterminate. This problem can be solved by applying l'Hopital's rule. The final result is
\begin{equation}
\label{4.17}
\left(\frac{\partial a_2}{\partial \xi^*}\right)_\text{s}= a_{2,\text{s}}\left(\zeta_\text{s}^* - \frac{\nu_{4|0^*}}{2} -q\xi_\text{s}^*\beta T_\text{s}^*-\frac{1-q}{2}\xi_\text{s}^*\right)^{-1}.
\end{equation}
Upon deriving Eq.\ \eqref{4.17}, use has been made of the identity
\begin{equation}
\label{4.18}
\frac{\partial}{\partial \xi^*}\left[\left( 2\beta T^* -1 \right) \xi^*\right]=\frac{2q}{1+q}\beta T^*-1.
\end{equation}
To obtain $\partial a_2/\partial \theta$ in the steady state, we derive first both sides of Eq.\ \eqref{4.16} with respect to $\theta$. The result is
\beq
\label{4.18.1}
(1+q)\left[ \left( 2\beta T^* -1 \right) \xi^* +\zeta^*\right]\xi^*\left(\frac{\partial^2 a_2}{\partial\xi^*\partial \theta}\right)
+2(1+q)\xi^{*\frac{1+2q}{1+q}}\left(\frac{\partial a_2}{\partial \xi^*}\right)
=2\left(\frac{\partial a_2}{\partial \theta}\right)\left(\zeta^*-\frac{1}{2}\nu_{4|0}^*-\xi^*\right).
\eeq
In the steady state, the first term of the left hand side of \eqref{4.18.1} vanishes and hence, one gets
\begin{equation}
\label{4.19}
\left(\frac{\partial a_2}{\partial \theta}\right)_{\text{s}}=(1+q)\frac{\xi^{*\frac{1+2q}{1+q}}}{\zeta_\text{s}^* -\frac{1}{2} \nu_{4|0^*}-\xi^*}\left(\frac{\partial a_2}{\partial \xi^*}\right)_\text{s},
\end{equation}
where use has been made of the result
\begin{equation}
\label{4.20}
\frac{\partial}{\partial \theta}(2\beta T^*)=\frac{\partial}{\partial \theta}
\frac{2\theta}{\xi^{*1/(1+q)}}=\frac{2}{\xi^{*1/(1+q)}}.
\end{equation}

\begin{figure}
{\includegraphics[width=0.49\columnwidth]{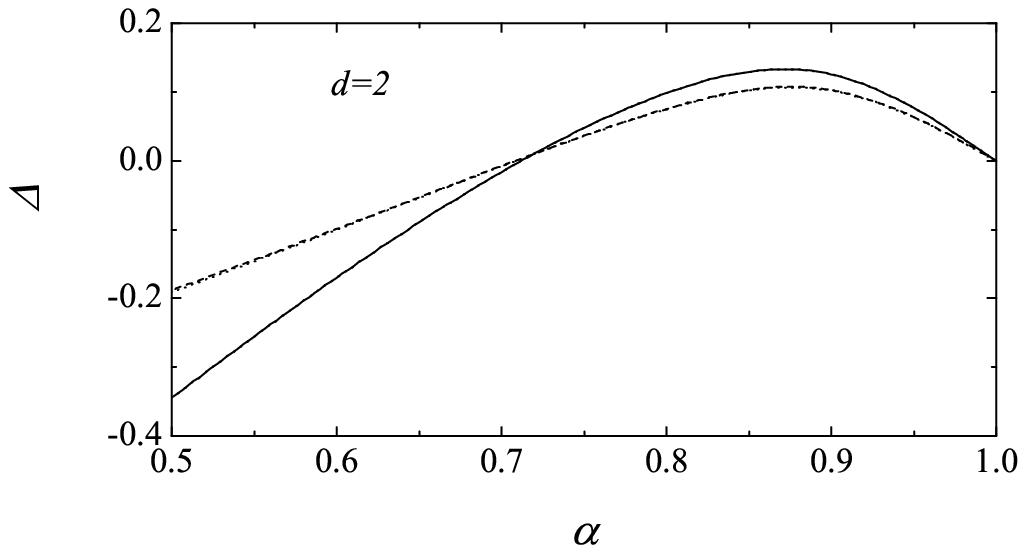}}
{\includegraphics[width=0.48\columnwidth]{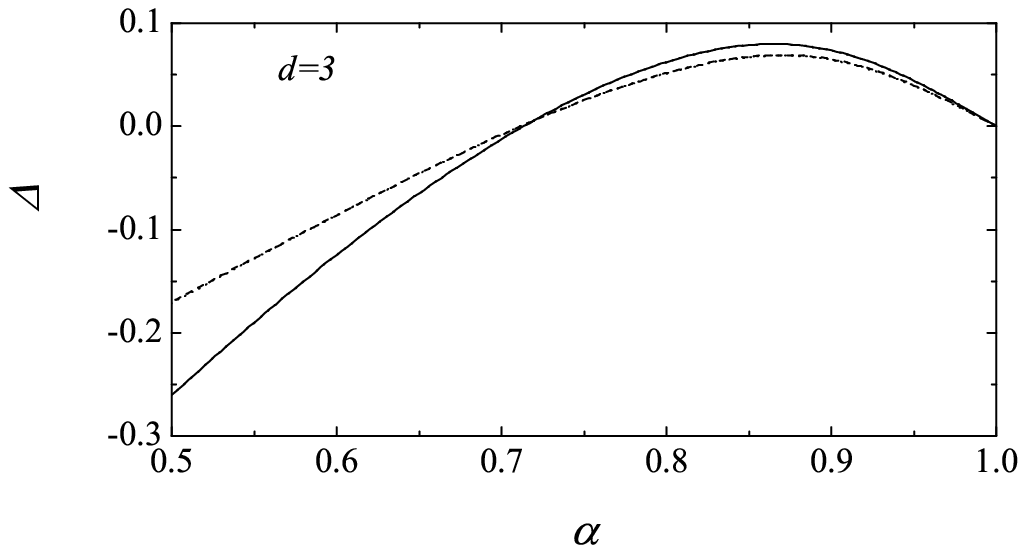}}
\caption{Plot of the derivative $\Delta\equiv \left(\frac{\partial a_2}{\partial \xi^*}\right)_\text{s}$ versus the coefficient of restitution $\alpha$ for the stochastic thermostat ($\xi_\text{s}^*=\zeta_\text{s}^*$) for disks ($d=2$) and spheres ($d=3$). The solid lines are the results given by Eq.\ \eqref{4.17} for $q=\frac{1}{2}$ while the dotted and dashed lines are the results obtained for IHS in Refs.\ \cite{GCV13} and \cite{GMT12}, respectively. Please note that the dotted and dashed lines overlap in the full interval represented here, meaning that both approaches in  \cite{GCV13} and \cite{GMT12} lead to identical results for this magnitude.
\label{fig2}}
\end{figure}

Figure \ref{fig2} shows the dependence of the derivative $\Delta\equiv \left(\frac{\partial a_2}{\partial \xi^*}\right)_\text{s}$ on the coefficient of restitution $\alpha$ when the gas is heated by the stochastic thermostat ($\beta=0$ and $\xi_\text{s}^*=\zeta_\text{s}^*$). The results obtained from Eq.\ \eqref{4.17} when $q=\frac{1}{2}$ are compared with those derived for IHS in Refs.\ \cite{GCV13} and \cite{GMT12} by using two different methods. First, it is quite apparent that the results obtained for IHS are practically indistinguishable, showing that the expressions of $\Delta$ obtained in Refs. \cite{GCV13} and \cite{GMT12} are consistent with each other. When comparing IHS and IMM, we observe that the discrepancies between both interaction models are very small for not too strong dissipation ($\al \gtrsim 0.6$), although they increase as the coefficient of restitution decreases.

\begin{figure}
{\includegraphics[width=0.4\columnwidth]{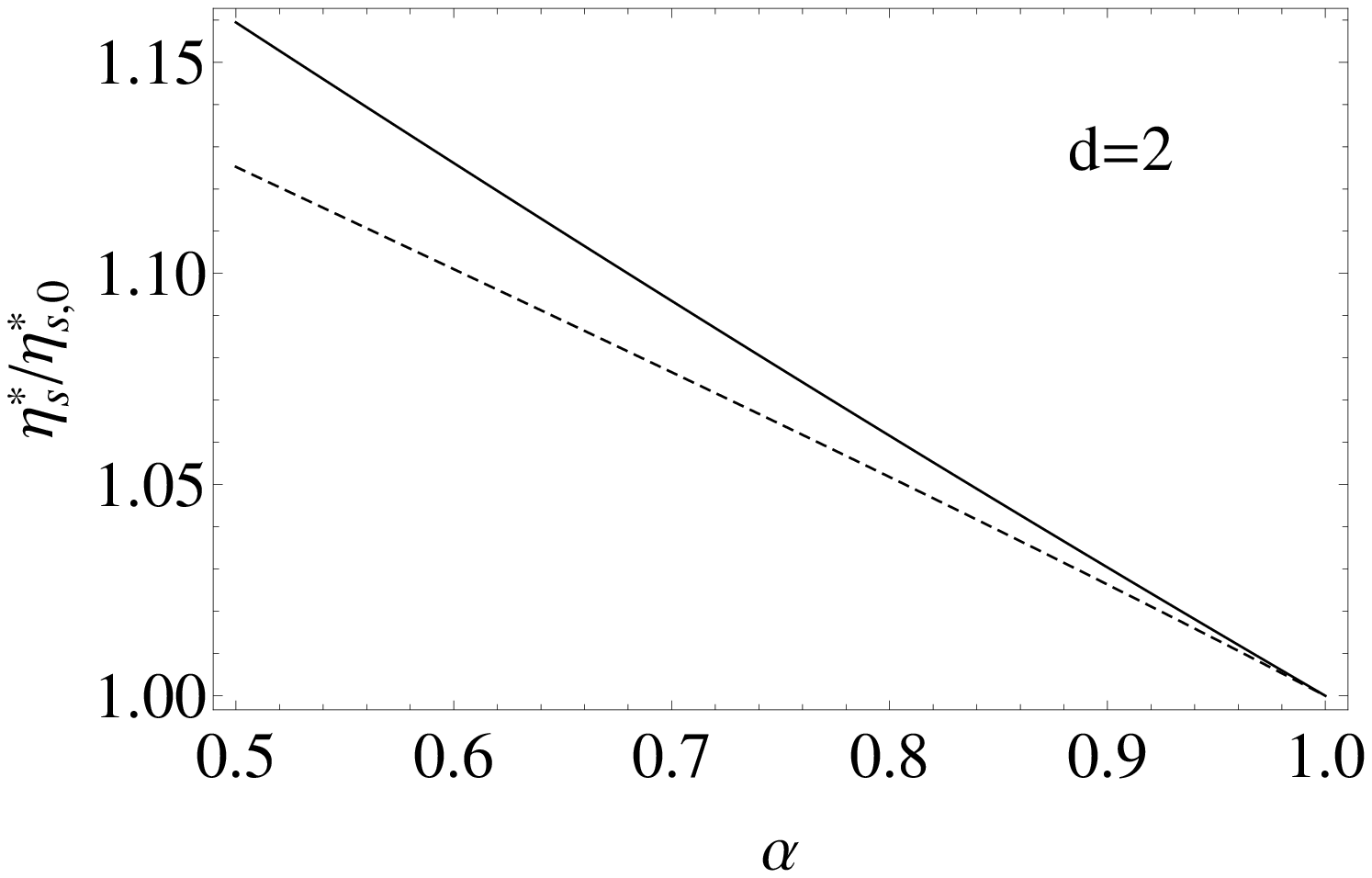}}
{\includegraphics[width=0.4\columnwidth]{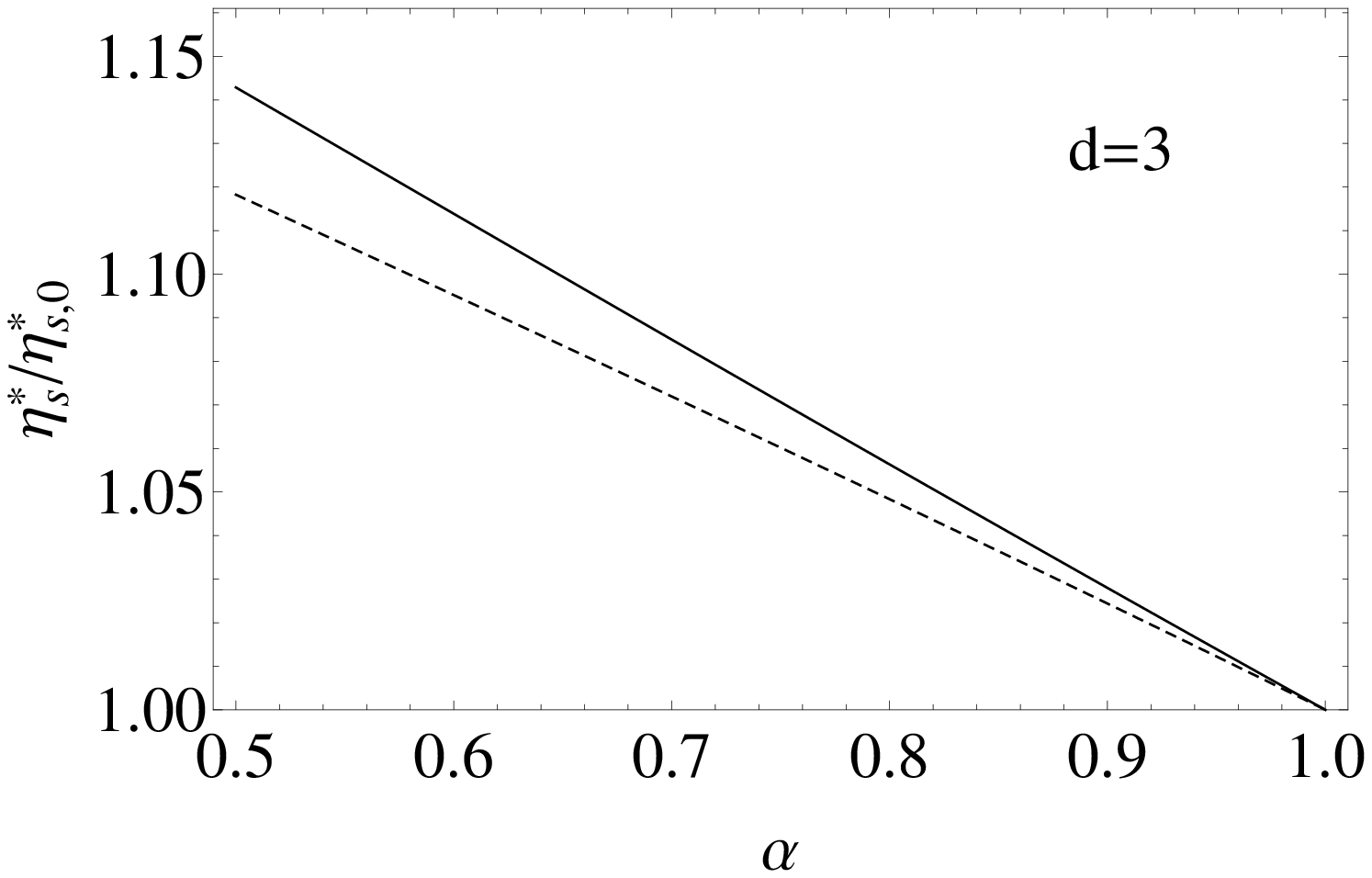}}
\caption{Plot of the reduced shear viscosity $\eta_\text{s}^*/\eta_\text{s,0}^*$ as a function of the coefficient of restitution $\al$ for $\beta=\frac{1}{2}$ in the case of a two- ($d=2$) and three-dimensional ($d=3$) system of IMM with $q=\frac{1}{2}$ (solid lines) and IHS (dashed lines). The value of the (reduced) noise strength is $\xi_\text{s}^*=1$.
\label{fig3}}
\end{figure}
\begin{figure}
{\includegraphics[width=0.4\columnwidth]{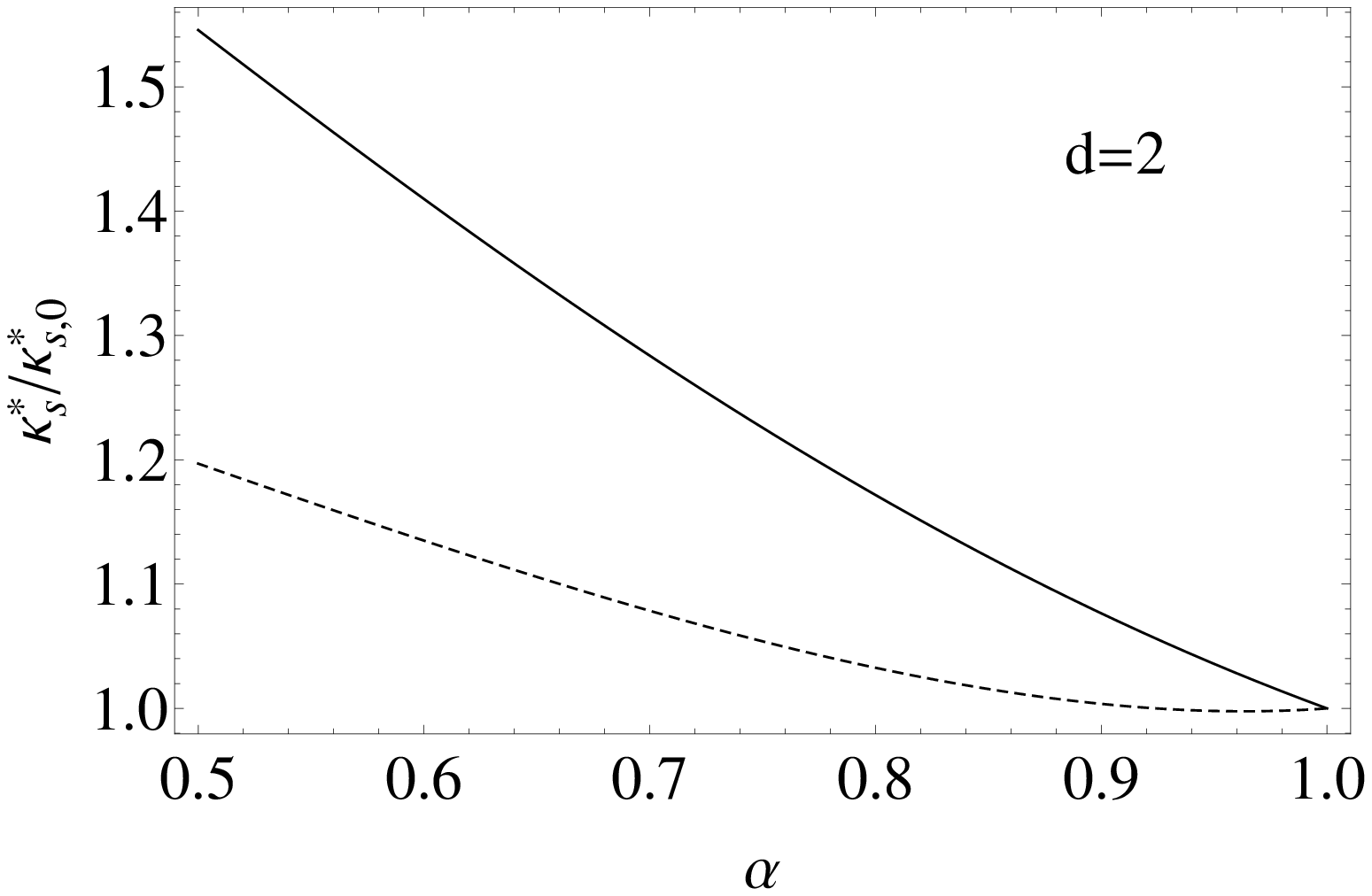}}
{\includegraphics[width=0.4\columnwidth]{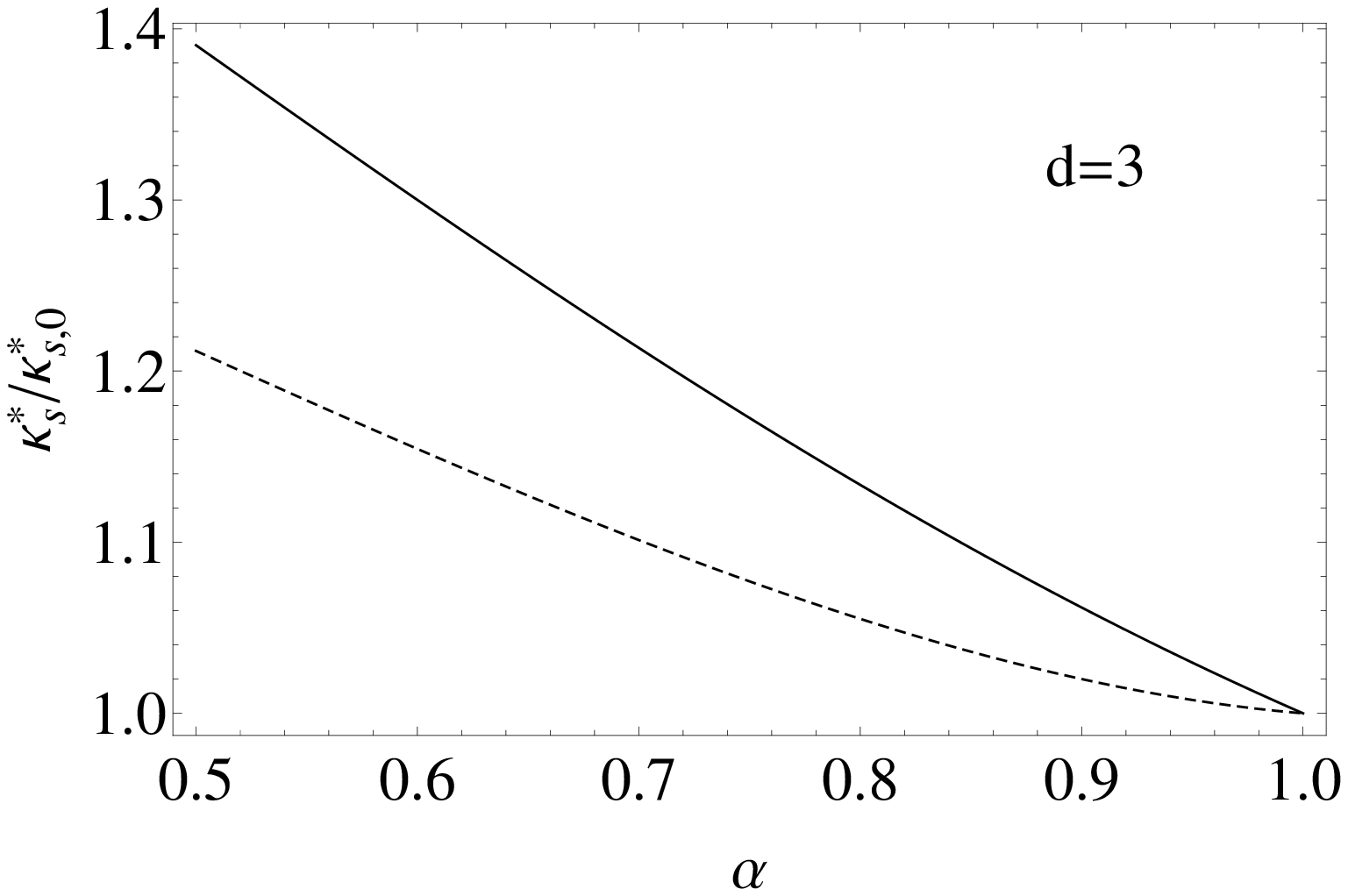}}
\caption{The same as in Fig.\ \ref{fig3} for the reduced thermal conductivity $\kappa_\text{s}^*/\kappa_\text{s,0}^*$.
\label{fig4}}
\end{figure}
\begin{figure}
{\includegraphics[width=0.4\columnwidth]{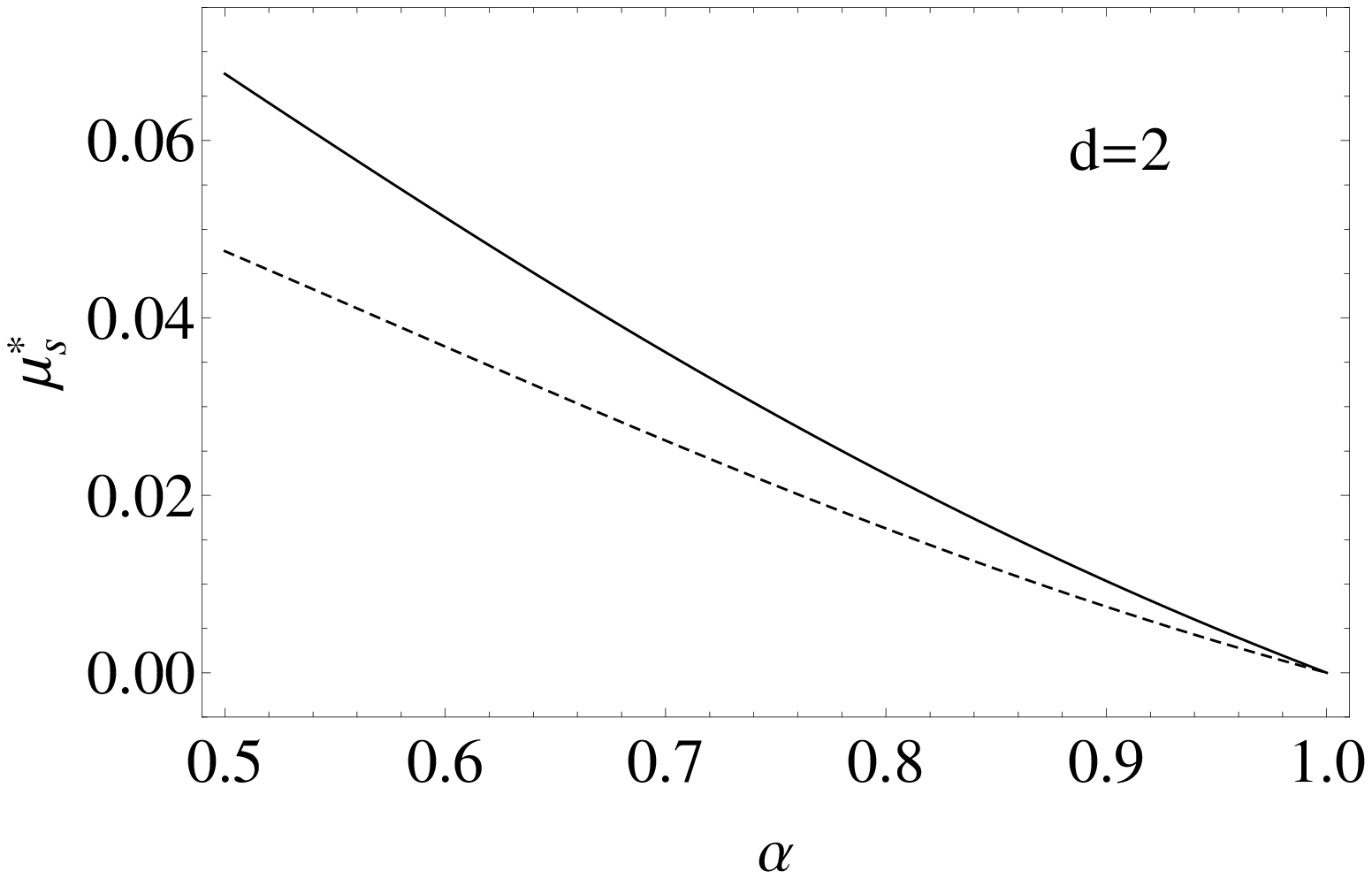}}
{\includegraphics[width=0.4\columnwidth]{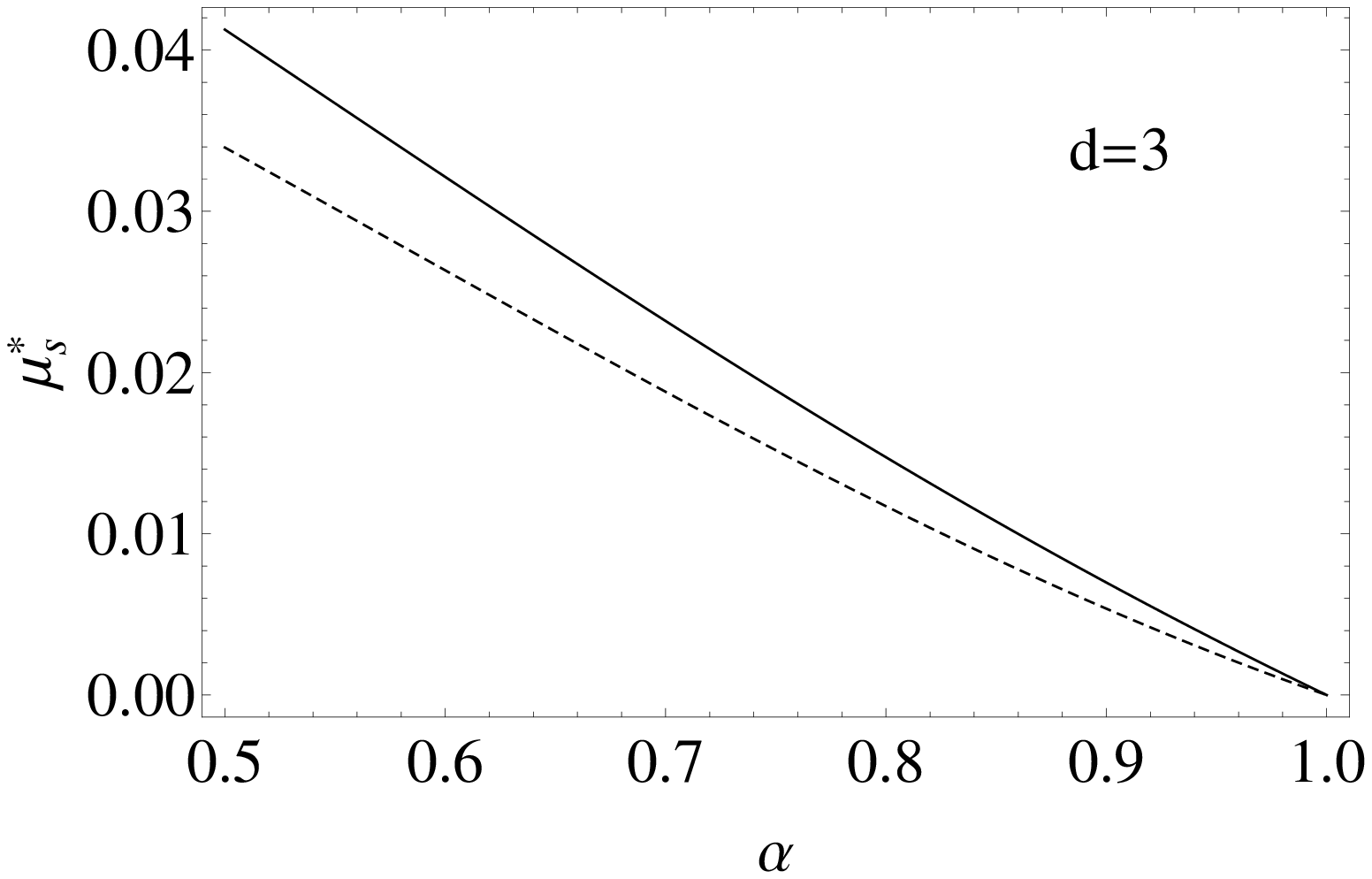}}
\caption{The same as in Fig.\ \ref{fig3} for the reduced coefficient $\mu_\text{s}^*= n\mu_\text{s}/\kappa_0 T$.
\label{fig5}}
\end{figure}
\begin{figure}
{\includegraphics[width=0.4\columnwidth]{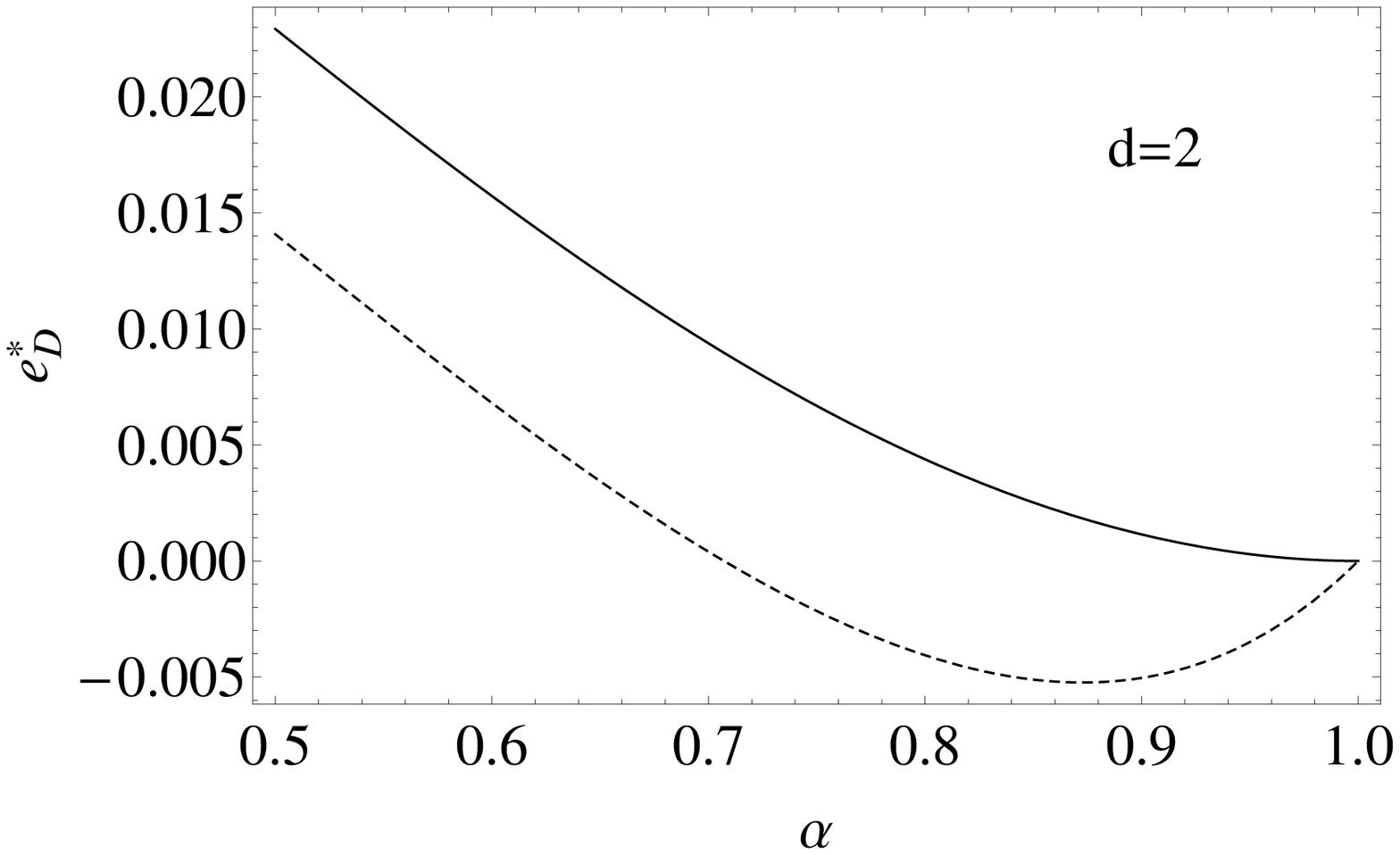}}
{\includegraphics[width=0.4\columnwidth]{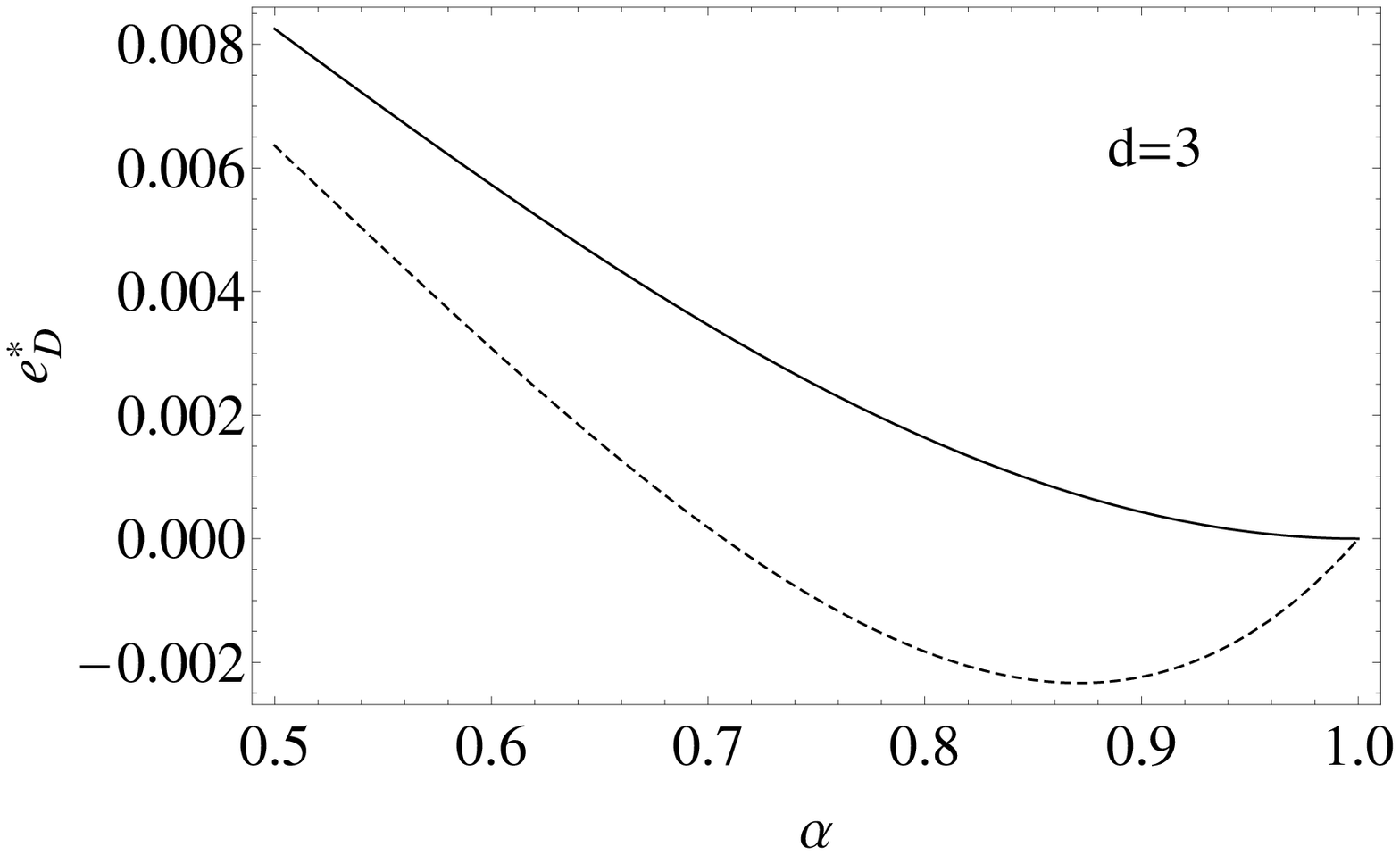}}
\caption{The same as in Fig.\ \ref{fig3} for the reduced coefficient $e_D^*$.
\label{fig6}}
\end{figure}

\subsection{Comparison with the Navier-Stokes transport coefficients of IHS in the steady state}
\label{sec6}

The Navier-Stokes transport coefficients of IHS have been derived in Ref.\ \cite{GCV13} in the first Sonine approximation for a driven granular dense gas. For the sake of completeness, the expressions of the reduced coefficients $\eta_\text{s}^*$, $\kappa_\text{s}^*$, $\mu_\text{s}^*$ and $e_D^*$ are listed in Appendix \ref{appB} for a low-density granular gas.

Figures \ref{fig3}--\ref{fig6} show the $\al$-dependence of the reduced transport coefficients $\eta_\text{s}^*/\eta_\text{s,0}^*$, $\kappa_\text{s}^*/\kappa_\text{s,0}^*$, $\mu_\text{s}^*$, and $e_D^*$, respectively, for $\xi_\text{s}^*=1$. Here, since we are mainly interested in analyzing the influence of dissipation on transport, the shear viscosity and thermal conductivity coefficients have been reduced with respect to their corresponding elastic values  $\eta_\text{s,0}^*$ and $\kappa_\text{s,0}^*$, respectively. Note that the coefficients $\mu_\text{s}^*$ and $e_D^*$ vanish for elastic collisions. In addition, we have taken $\beta=\frac{1}{2}$ and the Maxwell model with the power $q=\frac{1}{2}$. This latter choice is closer to inelastic hard spheres.

We observe that in general the qualitative dependence of the Navier-Stokes transport coefficients on dissipation of IHS is well captured by IMM. The shear viscosity (as expected because the same behavior is observed in analogous systems \cite{S03}) increases with inelasticity. However, this increase is faster for IMM. The (reduced) thermal conductivity of IHS presents a non-monotonic dependence with dissipation, since first it decreases as $\al$ decreases in the region of weak dissipation, reaches a minimum and then, the ratio $\kappa_\text{s}^*/\kappa_\text{s,0}^*$ increases with inelasticity. This behavior differs from the one observed for IMM where $\kappa_\text{s}^*/\kappa_\text{s,0}^*$ always increases with inelasticity. With respect to the new transport coefficient $\mu_\text{s}^*$ (not present for elastic collisions), both interaction models predict that this coefficient is much smaller than the thermal conductivity so that, the impact of the term $-\mu \nabla n$ on the heat flux $\mathbf{q}^{(1)}$ is much smaller than the Fourier's law term $-\kappa \nabla T$. Notice also that the quantitative differences between the Navier-Stokes transport coefficients of IMM and IHS  increase with inelasticity, especially in the two-dimensional case. However, and compared to freely cooling granular gases \cite{S03}, these quantitative differences between both models are much less important for driven systems. Therefore, we think the results in this paper are particularly useful also for studying the transport properties of the analogous IHS driven system.

\section{Discussion}
\label{sec6}

Calculation of the Navier-Stokes transport coefficients in driven granular gases from the Boltzmann equation for IHS is a quite difficult problem. In particular, one has to compute three different collision integrals to get the explicit forms of the Navier-Stokes transport coefficients. However, given that these integrals cannot be exactly evaluated, one usually considers the leading terms in a Sonine polynomial expansion of the velocity distribution function (first-Sonine approximation) to estimate them \cite{CC70}. In spite of the simplicity of this approach, the corresponding expressions of the Navier-Stokes transport coefficients compare in general quite well with computer simulations. On the other hand, it could be desirable to introduce interaction models more tractable analytically than IHS that were also capable of capturing the most important properties of the latter (at least within the domain of velocities near thermal velocity).

Based on the experience of elastic particles, a possible alternative that may overcome the technical difficulties embodied in the Boltzmann collision operator of IHS is to consider IMM. In the Boltzmann equation for IMM, the collision rate of the underlying system of IHS is replaced by an effective collision rate independent of the relative velocity of the two colliding particles. This property allows us to evaluate \emph{exactly} the velocity moments of the Boltzmann collision operator without the explicit knowledge of the velocity distribution function.

In this paper the expressions of the Navier-Stokes transport coefficients of an inelastic Maxwell gas driven by a stochastic bath with friction have been obtained. This type of thermostat (used in a number of works by other authors \cite{andrea}) is proposed to model the effect of the interstitial fluid on the dynamic properties of grains. As noted in the Introduction, the evaluation of the transport coefficients of IMM is an interesting problem by itself since it allows to understand in a clean way the influence of collisional dissipation on transport properties. In addition, the comparison between the \emph{exact} results for IMM with those obtained for IHS by using approximate analytical methods allows us to gauge the degree of reliability of IMM  for the description of granular flows. Here, we have accomplished this comparison with the results  for IHS derived by the authors in a recent work \cite{GCV13} by using the same type of thermostat.

The Navier-Stokes transport coefficients have been obtained by solving the Boltzmann equation for IMM by means of the Chapman-Enskog expansion up to first order in the spatial gradients. As noted in the previous work for IHS  \cite{GCV13}, collisional cooling cannot be necessarily balanced at all points in the system by the thermostat and/or external forces from the boundaries. As a consequence, the zeroth-order solution $f^{(0)}$ depends on time through its dependence on the granular temperature. The fact that  $\partial_t^{(0)}T \neq 0$ gives rise to conceptual and mathematical difficulties not present in previous works \cite{S03,GM02} where the parameters of the force were chosen to impose a steady temperature in the reference state $f^{(0)}$. In particular, we would need to solve numerically (which we have not done in the present work) a set of coupled first-order differential equations [see Eqs. \eqref{5.4}--\eqref{5.6}], in order to obtain the dependence of the transport coefficients on dissipation and the thermostat forces parameters. This technical difficulty is present even in the simplest Maxwell model where the collision frequency $\nu$ is independent of temperature $T$ [i.e., when $q=0$ in Eq.\ \eqref{2.4}]. Thus, we have considered the steady state conditions and have been able to obtain analytical expressions of all transport coefficients for this particular state. The steady state expressions are given by Eq.\ \eqref{5.14} for the (dimensionless) shear viscosity $\eta^*$, Eq.\ \eqref{5.15} for the (dimensionless) thermal conductivity $\kappa^*$, Eq.\ \eqref{5.16} for the coefficient $\mu^*$ and Eq.\ \eqref{5.17} for the first-order contribution $e_D^*$ to the fourth-cumulant. The three first coefficients provide the momentum and heat fluxes in the first order of the spatial gradients.

As in previous works \cite{S03,G07,SG07}, we choose the collision frequency $\nu$ appearing in the Boltzmann equation for IMM [see Eq.\ \eqref{2.2}] to reproduce the cooling rate $\zeta$ of IHS (evaluated in the Maxwellian approximation). With this choice, the comparison between IMM and IHS (see Figs.\ \ref{fig3}--\ref{fig6} for $d=2$ and 3) shows that IMM reproduce qualitatively well the trends observed for IHS, even for strong dissipation. On the other hand, at a more quantitative level, discrepancies between both interaction models increase with inelasticity, especially in the case of hard disks ($d=2$). In any case, the results found in this paper contrast with those obtained in the freely cooling case \cite{S03} where IMM and IHS exhibit much more significant differences. Thus, the reliability of IMM as a prototype model for granular flows can be considered more robust in driven states than in the case of undriven states. This conclusion agrees with the results derived in the case of the simple shear flow problem \cite{G03} and more complex shear-induced laminar flows \cite{SGV09}. In this context, the search for exact solutions for driven IMM, and comparison with computer simulations or experiments, can be considered as an interesting problem in the near future.

\acknowledgments
The present work has been supported by the Ministerio de
Educaci\'on y Ciencia (Spain) through grant No. FIS2010-16587, partially financed by
FEDER funds and by the Junta de Extremadura (Spain) through Grant No. GRU10158. The research of M. G. Chamorro has been supported by the predoctoral fellowship BES-2011-045869 from the Spanish Government (Spain).

\appendix
\section{First-order contributions to the fluxes}
\label{appA}

In this Appendix we determine the first-order contributions to the momentum and heat fluxes. Let us consider each flux separately. The first order contribution to the pressure tensor $P_{ij}^{(1)}$is defined by Eq.\ \eqref{5.1}. To obtain it, we multiply both sides of Eq.\ \eqref{4.21} by $m V_i V_j$ and integrate over $\mathbf{v}$. The result is
\beq
\label{a1}
\partial_t^{(0)}P_{ij}^{(1)}+\nu_{0|2}P_{ij}^{(1)}+\frac{2\gamma_\text{b}}{m} P_{ij}^{(1)}=-p
\left( \nabla_{i}U_{j}+\nabla_{j}U_{i}-\frac{2}{d}\delta_{ij}\nabla \cdot
\mathbf{U} \right).
\eeq
Upon writing Eq.\ \eqref{a1}, use has been made of the result
\beq
\label{a2}
\int \dd \mathbf{v}\; m V_i V_j {\cal L}f^{(1)}=\nu_{0|2} P_{ij}^{(1)},
\eeq
where $\nu_{0|2}$ is given by Eq.\ \eqref{2.17}. The solution to Eq.\ \eqref{a1} can be written in the form \eqref{5.2.1}, where the shear viscosity coefficient $\eta$ obeys the time dependent equation
\beq
\label{a4}
\partial_t^{(0)}\eta+\left(\nu_{0|2}+\frac{2\gamma_\text{b}}{m}\right)\eta=p.
\eeq
The shear viscosity can be written in the form \eqref{5.3} where $\eta^*$ is a dimensionless function of the reduced noise strength $\xi^*$ (or the reduced drag parameter $\gamma^*$  through Eq.\ \eqref{4.11}) and the coefficient of restitution $\al$. Thus,
\beq
\label{a5}
\partial_t^{(0)}\eta=(T\partial_T \eta)(\partial_t^{(0)} \ln T)=\Lambda T\partial_T (\eta_0 \eta^*)=
\Lambda \left[(1-q)\eta-(1+q)\eta_0\xi^*\frac{\partial \eta^*}{\partial \xi^*}\right],
\eeq
where
\beq
\label{a6}
\Lambda \equiv \frac{m\xi_b^2}{T}-\frac{2\gamma_b}{m} -\zeta.
\eeq
Equation \eqref{5.4} for $\eta^*$ can be easily obtained when one takes into account the relation \eqref{a5} in Eq.\ \eqref{a4}.

The first order contribution to the heat flux is defined by Eq.\ \eqref{5.2}. As in the case of the pressure tensor, to obtain $\mathbf{q}^{(1)}$ we multiply both sides of Eq.\ \eqref{4.21} by $\frac{m}{2} V^2 \mathbf{V}$ and integrate over $\mathbf{v}$. After some algebra, one gets
\beqa
\label{a9}
&&\partial_t^{(0)}\mathbf{q}^{(1)}+\left(\nu_{2|1}+\frac{3\gamma_\text{b}}{m}\right)\mathbf{q}^{(1)}=
-\frac{d+2}{2}\frac{p}{m}\left(1+2a_2-(1+q)\xi^*\frac{\partial a_2}{\partial \xi^*}\right)\nabla T \nonumber\\
& &-\frac{d+2}{2}\frac{T^2}{m}\left(a_2-\frac{\theta}{1+q}\frac{\partial a_2}{\partial \theta}-\xi^*\frac{\partial a_2}{\partial \xi^*}\right)\nabla n.
\eeqa
Upon writing Eq.\ \eqref{a9}, the following results have been used:
\beq
\label{a10.1}
\int \dd \mathbf{v}\; \frac{m}{2} V^2 \mathbf{V} {\cal L}f^{(1)}=\nu_{2|1} \mathbf{q}^{(1)},
\eeq
\beqa
\label{a10}
\int\; \dd \mathbf{v}\; \frac{m}{2}V^2 V_i A_j(\mathbf{V})&=&
-\frac{d+2}{2}\frac{p T}{m}\delta_{ij}\left(1+2a_2+T\partial_T a_2\right) \nonumber\\
&=& -\frac{d+2}{2}\frac{p T}{m}\delta_{ij}\left(1+2a_2-
(1+q)\xi^*\frac{\partial a_2}{\partial \xi^*}\right),
\eeqa
\beqa
\label{a11}
\int\; \dd \mathbf{v}\; \frac{m}{2}V^2 V_i B_j(\mathbf{V})&=&-\frac{d+2}{2}\frac{p T}{m}\delta_{ij}\left(a_2+n\partial_n a_2\right) \nonumber\\
&=&-\frac{d+2}{2}\frac{p T}{m}\delta_{ij}\left(a_2-\frac{\theta}{1+q}\frac{\partial a_2}{\partial \theta}-\xi^*\frac{\partial a_2}{\partial \xi^*}\right).
\eeqa
In Eq.\ \eqref{a10.1}, $\nu_{2|1}$ is defined by Eq.\ \eqref{2.18}. The solution to Eq.\ \eqref{a9} is given by Eq.\ \eqref{5.2.2}, where $\kappa$ is the thermal conductivity coefficient and $\mu$ is a new coefficient not present for elastic collisions. The Navier-Stokes transport coefficients $\kappa$ and $\mu$ can be written in the form \eqref{5.3}, where the (reduced) coefficients $\kappa^*$ and $\mu^*$ depend on $T$ through their dependence on $\xi^*$:
\beq
\label{a14}
\partial_t^{(0)}\kappa=(T\partial_T \kappa)(\partial_t^{(0)} \ln T)=\Lambda T\partial_T (\kappa_0 \kappa^*)=
\Lambda \left[(1-q)\kappa-(1+q)\kappa_0\xi^*\frac{\partial \kappa^*}{\partial \xi^*}\right],
\eeq
\beq
\label{a15}
\partial_t^{(0)}\mu=(T\partial_T \mu)(\partial_t^{(0)} \ln T)=\Lambda T\partial_T \left(\frac{\kappa_0 T}{n} \mu^*\right)=\Lambda \left[(2-q)\mu-(1+q)\frac{\kappa_0 T}{n} \xi^*\frac{\partial \mu^*}{\partial \xi^*}\right].
\eeq
Moreover, there are also contributions to Eq.\ \eqref{a9} coming from the term
\beq
\label{a16}
\nabla \partial_t^{(0)} T=\left(\Lambda-\frac{m\xi_\text{b}^2}{T}-q \zeta\right) \nabla T-\frac{T \zeta}{n}\nabla n.
\eeq
The corresponding differential equations for $\kappa^*$ and $\mu^*$ can be obtained when one takes into account the constitutive form \eqref{5.2} and the relations  \eqref{a14}--\eqref{a16} in Eq.\ \eqref{a9}. These equations are given by Eq.\ \eqref{5.5} for $\kappa^*$ and Eq.\ \eqref{5.6} for $\mu^*$.


We consider finally the isotropic fourth degree moment \eqref{5.8}. Since $e_\text{D}$ is a scalar, it can be only coupled to the divergence of flow velocity $\nabla \cdot \mathbf{U}$:
\beq
\label{a20}
e_D=e_D^* \nu^{-1} \nabla \cdot \mathbf{U}.
\eeq
In order to determine the (reduced) coefficient $e_D^*$, we multiply both sides of Eq.\ \eqref{4.1} by $V^4$ and integrate over velocity. After some algebra one arrives to Eq.\ \eqref{5.10} where use has been made of the partial result
\beq
\label{a23}
\int\; \dd \mathbf{v}\; V^4 D(\mathbf{V})=d(d+2)\frac{nT^2}{m^2}\left(\frac{2(1+q)+d}{d}\xi^*\frac{\partial a_2}{\partial \xi^*}+\frac{\theta}{1+q}\frac{\partial a_2}{\partial \theta}\right).
\eeq

\section{Navier-Stokes transport coefficients for IHS in the steady state}
\label{appB}

In this Appendix we display the explicit expressions of the Navier-Stokes transport coefficients obtained in Ref.\ \cite{GCV13} for a moderately dense gas by considering the leading terms in a Sonine polynomial expansion. In the low-density limit, the forms of the dimensionless coefficients $\eta_\text{s}^*$, $\kappa_\text{s}^*$, and $\mu_\text{s}^*$ for IHS in the steady state are given, respectively, by
\beq
\label{b1}
\eta_\text{s}^*=\frac{2}{d+2}\frac{1}{\nu_\eta^*+2\gamma_\text{s}^*},
\eeq
\beq
\label{b2}
\kappa_\text{s}^*=\frac{2(d-1)}{d(d+2)}\frac{1+2 a_{2,\text{s}}-\frac{3}{2}\xi_\text{s}^*\Delta_\xi}{\nu_\kappa^*+
\frac{\xi_\text{s}^*}{2}\left[1+\frac{9}{32d}(1-\al^2)\Delta_\xi\right]-2\zeta_\text{s}^*},
\eeq
\beq
\label{b3}
\mu_\text{s}^*=\frac{
\kappa_\text{s}^*\left[\zeta_\text{s}^*-\frac{3(1-\al^2)}{32d}\left(\theta_\text{s}\Delta_\theta+
\xi_\text{s}^*\Delta_\xi\right)\right]
+\frac{2(d-1)}{d(d+2)}\left[
a_{2,\text{s}}-\theta_\text{s} \Delta_\theta-
\xi_\text{s}^*\Delta_\xi\right]}
{\nu_{\kappa}^*+3\gamma_\text{s}^*},
\eeq
where
\beq
\label{b4}
\nu_\eta^*=\frac{3-3\al+2d}{2d(d+2)}(1+\al)\left(1+\frac{7}{16}a_{2,\text{s}}\right),
\eeq
\beq
\label{b5}
\nu_\kappa^*=\frac{2}{d(d+2)}(1+\al)\left[\frac{d-1}{2}+\frac{3}{16}(d+8)(1-\alpha)
+\frac{296+217d-3(160+11d)\alpha}{256}a_{2,\text{s}}\right],
\eeq
\beq
\label{b6}
\zeta_\text{s}^*=\frac{1-\al^2}{2d}\left(1+\frac{3}{16}a_{2,\text{s}}\right),
\eeq
and
\begin{equation}
\label{b7}
a_{2,\text{s}}=\frac{16(1-\alpha)(1-2\alpha^2)}{9+24d-\alpha(41-8d)+30(1-\alpha)\alpha^2
+\frac{64d(d+2)}{1+\alpha}\xi_\text{s}^*}.
\end{equation}
In addition, the quantities $\Delta_\xi$ and $\Delta_\theta$ are related to the derivatives $(\partial a_2/\partial \xi^*)_\text{s}$ and $(\partial a_2/\partial \theta)_\text{s}$, respectively. The derivative $(\partial a_2/\partial \xi^*)_\text{s}$ obeys the quadratic equation (A6) of Ref.\ \cite{GCV13}. However, given that the magnitude of this derivative is in general quite small, one can neglect the nonlinear term  $(\partial a_2/\partial \xi^*)^2$ in this quadratic equation and get an explicit expression for this derivative. In this approximation, the quantities $\Delta_\xi$ and $\Delta_\theta$ can be written as
\beq
\label{b8}
\Delta_\xi=\frac{ a_{2,\text{s}}}{\frac{19}{32d}(1-\al^2)-\frac{1+\left(\frac{2^{5/2}}{d+2}\right)^{2/3}\theta_\text{s}\xi_\text{s}^{*-2/3} }{4}\xi_\text{s}^*-\frac{1-\al^2}{2d(d+2)}
\left[\frac{3}{32}(10d+39+10\al^2)+\frac{d-1}{1-\al}\right]}.
\eeq
\beq
\label{b9}
\Delta_\theta=\frac{\left(\frac{2^{5}}{(d+2)^2}\right)^{1/3}\xi_\text{s}^{*4/3}\Delta_\xi}{\frac{3}{16d}(1-\al^2)\left(1+
a_{2,\text{s}}-\frac{3}{4}\xi_\text{s}^*\Delta_\xi\right)+2(\zeta_\text{s}^*-\xi_\text{s}^*)-\frac{1-\al^2}{d(d+2)}
\left[\frac{3}{32}(10d+39+10\al^2)+\frac{d-1}{1-\al}\right]},
\eeq
where
\beq
\label{b10}
\theta_\text{s}=\frac{\xi_\text{s}^*-\zeta_\text{s}^*}{2}\xi_\text{s}^{*1/3}.
\eeq

Finally, the coefficient $e_D^*$ is
\beq
\label{b11}
e_D^*=-\frac{\frac{d+3}{2d}\xi_\text{s}^*\Delta_\xi
+\frac{1}{2}\theta_\text{s}\Delta_\theta}{\nu_{\gamma}^*+4\gamma_\text{s}^*},
\eeq
where
\beq
\label{b12}
\nu_\gamma^*=-\frac{2}{96(d+2)}(1+\al)\left[30\alpha^3-30\alpha^2+(105+24 d) \alpha-56d-73\right].
\eeq

\end{document}